\documentclass[prd,preprint,superscriptaddress,preprintnumbers,eqsecnum,showpacs,nofootinbib,nobibnotes,noeprint]{revtex4-1}
\usepackage{amsfonts,amsmath,mathbbol,amssymb,bm,natbib}
\usepackage[table,dvipsnames]{xcolor}
\usepackage{graphicx} 
\usepackage{color}
\usepackage{stackengine}
\usepackage{scalerel}
\usepackage{mathtools}
\usepackage[linkcolor=blue,citecolor=blue,urlcolor=blue,colorlinks=true]{hyperref}
\definecolor{refkey}{rgb}{0.39,0.58,1}
\definecolor{labeled}{rgb}{1,0,0}
\usepackage{multirow}
\usepackage{mathrsfs}
\usepackage{array}
\usepackage{tabularx}
\usepackage{float}
\usepackage{physics}
\usepackage{comment}
\usepackage[export]{adjustbox}
\usepackage{slashed}

\def\ie{{\it i.e.}, }
\newcommand{\be}{\begin{equation}}
\newcommand{\bea}{\begin{eqnarray}}
\newcommand{\ee}{\end{equation}}
\newcommand{\eea}{\end{eqnarray}}

\def\s#1{{\scriptscriptstyle #1}}

\def\eq#1{Eq.~(\ref{#1})}
\def\1eq#1{Eq.~(\ref{#1})}

\def\2eqs#1#2{Eqs.~(\ref{#1}) and~(\ref{#2})}
\def\3eqs#1#2#3{Eqs.~(\ref{#1}),~(\ref{#2}) and~(\ref{#3})}
\def\6eqs#1#2#3#4#5#6{Eqs.~(\ref{#1}),~(\ref{#2}),~(\ref{#3}),~(\ref{#4}),~(\ref{#5}) and~(\ref{#6})} 
\def\fig#1{Fig.~\ref{#1}}
\def\eg{{\it e.g.}, }
\def\eg{{\it e.g.}}

\newcommand{\gh}{\widehat{\Gamma}}
\newcommand{\gt}{\widetilde{\Gamma}}
\newcommand{\gb}{\widehat{\bm{\Gamma}}}
\newcommand{\ph}{\widehat{\Pi}}

\newcommand{\yt}{\widetilde{R}}
\newcommand{\yy}{R}

\newcommand{\msym}{u}

\newcommand{\bbb}{BBB}
\newcommand{\bqq}{BQQ}
\newcommand{\bbqq}{BBQQ}
\newcommand{\bqqq}{BQQQ}
\newcommand{\bbqqq}{BBQQQ}
\newcommand{\bbcc}{${\rm BB\bar{c}c}$}
\newcommand{\bqcc}{${\rm BQ\bar{c}c}$}
\newcommand{\bcc}{${\rm B\Bar{c}c}$}
\newcommand{\bbqcc}{${\rm BBQ\Bar{c}c}$}

\begin{document}

\title{Patterns of gauge symmetry in the background field method}

\author{A.~C.~Aguilar}
\affiliation{\mbox{University of Campinas - UNICAMP, Institute of Physics ``Gleb Wataghin'',} 
13083-859 Campinas, S\~{a}o Paulo, Brazil.}

\author{M.~N. Ferreira}
\affiliation{\mbox{Department of Theoretical Physics and IFIC,} \\ University of Valencia and CSIC, E-46100, Valencia, Spain.}

\author{D. Iba\~nez}
\affiliation{\mbox{University Centre EDEM}, Muelle de la Aduana, La Marina de Valencia, 46024, Valencia, Spain.}

\author{B.~M. Oliveira}
\affiliation{\mbox{University of Campinas - UNICAMP, Institute of Physics ``Gleb Wataghin'',} 13083-859 Campinas, S\~{a}o Paulo, Brazil.}

\author{J.~Papavassiliou}
\affiliation{\mbox{Department of Theoretical Physics and IFIC,} \\ University of Valencia and CSIC, E-46100, Valencia, Spain.}

%\pacs{
%12.38.Aw,  % General properties of QCD (dynamics, confinement, etc)
%12.38.Lg, % Other nonperturbative calculations
%14.70.Dj %Gluons
%}

\begin{abstract}

The correlation functions of Yang-Mills theories formulated in 
the background field method  
satisfy linear Slavnov-Taylor identities,  which are naive generalizations of simple 
tree level  relations, with no deformations originating from the 
ghost-sector of the theory. In recent years, 
a stronger version of these identities has been found to hold 
at the level of the background gluon self-energy, whose transversality 
is enforced separately for each special block of diagrams 
contributing to the gluon Schwinger-Dyson equation. 
In the present work we demonstrate by means of explicit calculations 
that the same distinct realization of the Slavnov-Taylor identity 
persists in the case of the background three-gluon vertex.   
The analysis is carried out at the level of the exact 
Schwinger-Dyson equation for this vertex, with no truncations or 
simplifying assumptions. The demonstration entails   
the contraction of individual 
vertex diagrams by the relevant momentum, which activates   
Slavnov-Taylor identities of vertices and multi-particle 
kernels nested inside these 
graphs; the final result emerges by virtue of a multitude of extensive cancellations, without the need of 
performing explicit integrations. 
In addition, we  point out that background Ward identities 
amount to \mbox{replacing} derivatives of propagators by zero-momentum background-gluon insertions, 
in exact \mbox{analogy} to standard properties of Abelian 
gauge theories. Finally, certain  potential 
applications of these results are briefly discussed.

\end{abstract}

\maketitle

\section{Introduction}
 
 In recent years, the systematic exploration of Green's (correlation) functions has afforded  
important insights on the nonperturbative properties of non-Abelian gauge theories, such as
pure Yang-Mills theories and Quantum Chromodynamics~\cite{Roberts:1994dr,Alkofer:2000wg,Fischer:2006ub,Roberts:2007ji,Binosi:2009qm,Binosi:2014aea,Cloet:2013jya,Aguilar:2015bud,Binosi:2016rxz,Binosi:2016nme,Huber:2018ned,Papavassiliou:2022wrb}. This ongoing scrutiny relies on
continuum studies based on nonperturbative functional methods~\cite{Maris:1997tm, Maris:2003vk, Braun:2007bx,Eichmann:2008ef, Cloet:2008re,Boucaud:2008ky,Eichmann:2009qa,Fischer:2008uz,Boucaud:2008ky,Dudal:2008sp,RodriguezQuintero:2010wy,Tissier:2010ts,Pennington:2011xs,Huber:2012zj,Cloet:2013jya,Fister:2013bh,Cyrol:2014kca,Boucaud:2008ky,Pawlowski:2003hq,Pawlowski:2005xe,Cyrol:2017ewj,Cyrol:2018xeq,Corell:2018yil, Gao:2017uox, Blaizot:2021ikl,Roberts:2020hiw,Roberts:2020hiw,Horak:2021pfr,Gao:2021wun}
carried out almost exclusively in the linear
covariant ($R_\xi$) gauges, where the Landau gauge is the preferred choice,
and on lattice simulations performed in the same gauge~\cite{Cucchieri:2006tf,Cucchieri:2007md,Bogolubsky:2007ud,Bogolubsky:2009dc,Cucchieri:2008qm,Cucchieri:2009zt, Oliveira:2009eh,Oliveira:2010xc,Maas:2011se,Ayala:2012pb,Oliveira:2012eh,Athenodorou:2016oyh,Duarte:2016ieu,Boucaud:2017ksi,Boucaud:2018xup,Aguilar:2021okw}.
However, the background field method (BFM) \cite{DeWitt:1967ub,Honerkamp:1972fd,Kallosh:1974yh,Kluberg-Stern:1974nmx,Arefeva:1974jv,Abbott:1980hw,Weinberg:1980wa,Abbott:1981ke,Shore:1981mj,Abbott:1983zw} has also been 
employed in several occasions, furnishing
useful vantage points, and exposing key properties of the theory that are normally distorted
by standard quantization procedures~\cite{Aguilar:2006gr,Aguilar:2008xm,Aguilar:2015bud,Papavassiliou:2022wrb}.

The BFM is a powerful framework that enables the 
implementation of  the gauge-fixing procedure necessary for
quantizing gauge theories without losing explicit gauge invariance, in contradistinction 
to the conventional quantization schemes~\cite{DeWitt:1967ub,Honerkamp:1972fd,Kallosh:1974yh,Kluberg-Stern:1974nmx,Arefeva:1974jv,Abbott:1980hw,Weinberg:1980wa,Abbott:1981ke,Shore:1981mj,Abbott:1983zw}. 
The starting point of the BFM is the splitting of the gauge field $A_{\mu}^{a}$ appearing in the classical action of the theory   
according to \mbox{$A_{\mu}^{a} = B_{\mu}^{a} +Q_{\mu}^{a}$}, where $B_{\mu}^{a}$ and $Q_{\mu}^{a}$ are the background and quantum (fluctuating) fields, respectively. The 
quantum field is the variable of integration in the generating functional $Z(J)$, and external sources are coupled only to it, as $J\cdot Q$. The background field does not enter in loops; it couples externally to Feynman diagrams, connecting them with the asymptotic states to form S-matrix elements. 
Then, by virtue of a special gauge-fixing condition, the resulting action (with the corresponding ghost terms included)
is no longer invariant under transformations of the quantum field, but retains its invariance intact 
with respect to the background field~\cite{Abbott:1981ke,Binosi:2009qm}.

A key consequence of the background gauge invariance of the action 
is that Green's functions involving the $B_{\mu}^{a}$ field satisfy 
ghost-free Slavnov-Taylor identities (STIs), akin to the 
Takahashi identity known from QED: the STIs are straightforward generalization of tree level relations,
receiving no ghost-related contributions after the inclusion of quantum corrections. 
Instead, in the standard STIs~\cite{Taylor:1971ff,Slavnov:1972fg} of the $R_\xi$ gauges~\cite{Fujikawa:1972fe}, starting already at one-loop, 
the ghost sector
modifies these tree level relations non-trivially.
To fix the ideas with a simple example,  the quark-gluon 
vertex with either a $B$ or a $Q$ gluon 
satisfies 
at tree level the simple identity  (suppressing color)
$q^{\mu}\gamma_\mu = \slashed{q} =  (-\slashed{p}-m) - (\slashed{r}-m) =  S_0^{-1}(-\slashed{p}) -S_0^{-1}(\slashed{r})$, 
where $S_0$ is the tree level version of the quark propagator $S$. 
In the case of a $B$ gluon, the all-order STI is obtained from the 
above tree level identity by simply substituting $S_0 \to S$, namely 
\mbox{$q^{\mu}\widehat\Gamma_{\mu}(q,r,p) = S^{-1}(-\slashed{p}) -S^{-1}(\slashed{r})$}.
Instead, in the case of a $Q$ gluon ($R_\xi$ gauges),
the STI gets modified by quantum corrections~\cite{Taylor:1971ff,Slavnov:1972fg}, which induce a dependence 
on the ghost dressing function
and the quark-ghost kernel~\cite{Marciano:1977su,Davydychev:2000rt,Aguilar:2010cn,Aguilar:2016lbe}.

As was pointed in earlier studies,  
the STI satisfied by the 
background-gluon self-energy, $\widehat{\Pi}_{\mu\nu}(q)$, namely the standard transversality condition 
\mbox{$q^{\mu} \widehat{\Pi}_{\mu\nu}(q)= 0$}, is implemented in a very special way, 
which has been denominated ``block-wise''~\cite{Aguilar:2006gr,Aguilar:2008xm,Aguilar:2015bud}. 
Specifically, 
the diagrammatic representation of the SDE governing $\widehat{\Pi}_{\mu\nu}(q)$ 
is composed by four distinct blocks,  namely 
one- and two-loop diagrams containing 
only gluons, 
and one- and two-loop diagrams containing 
 ghost fields, as shown in 
\fig{fig_bbsde}. 
The block-wise realization of the STI in this case is the simple statement 
that  the transversality 
of $\widehat{\Pi}_{\mu\nu}(q)$ 
is enforced independently for each of the four blocks. 
This is in sharp contrast to what happens
in the $R_\xi$ gauges, where, already at the one-loop perturbative level, 
it is only the sum of gluon and ghost diagrams 
that is transverse~\cite{Itzykson:1980rh,Peskin:1995ev}. The proof of this property is particularly simple; 
it proceeds by contracting the various diagrams by 
$q^{\mu}$ from the side of the fully dressed vertices,
thus triggering the corresponding naive STIs of the BFM. 
It is important to emphasize that the proof holds for any value of the gauge-fixing 
parameter $\xi_{\s Q}$ used to define the propagators of the $Q$-type gluons entering in the various loops.

The basic question that arises naturally in this context is whether 
the special block-wise realization of the STI described above is particular 
to the two-point 
function, or if it is a common feature of all  
Green's functions containing only $B$ gluons. 
In the present work we take a first step in the exploration of this issue, 
and demonstrate that the same pattern persists in the STI 
of the BFM vertex with three incoming background gluons, to be denoted by $\gh_{\alpha\mu\nu}(q,r,p)$.
This Abelian STI 
relates the contraction $q^\mu \gh_{\alpha\mu\nu}(q,r,p)$
to the difference \mbox{$\widehat{\Pi}_{\mu\nu}(r) - \widehat{\Pi}_{\mu\nu}(p)$}~\cite{Cornwall:1989gv,Aguilar:2006gr,Binosi:2009qm}.
It turns out that the diagrams comprising the SDE of $\gh_{\alpha\mu\nu}(q,r,p)$ 
may also be classified into four subsets in a way completely analogous 
to the case of $\widehat{\Pi}_{\mu\nu}(q)$.
Then, the 
contraction of each subset by $q^\mu$
generates the difference of the {\it corresponding} subsets 
of $\widehat{\Pi}_{\mu\nu}$, confirming the block-wise realization of this STI. 
We emphasize that, as in the case of the gluon self-energy, this property is completely 
$\xi_{\s Q}$-independent.  

An additional noteworthy aspect of the BFM Green's functions is the  
Ward identities (WIs) they satisfy, namely the relations that emerge 
when the momentum that triggers the STIs is taken to vanish. 
For example, in the case of the $\widehat\Gamma_{\mu}(q,r,p)$ mentioned above, 
a Taylor expansion of the STI around $q=0$ and subsequent matching 
of terms linear in $q$ yields the relation 
$\widehat\Gamma_{\mu}(0,-p,p) = \partial S^{-1}(\slashed{p})/\partial p^{\mu}$, which is 
the precise equivalent of the text-book WI known from QED, 
relating the photon-electron vertex with the electron propagator~\cite{Itzykson:1980rh,Peskin:1995ev}. 
In fact, exactly as happens in QED, this WI admits 
a simple diagrammatic interpretation: the derivative of the inverse 
quark propagator may be depicted 
as the insertion of a background gluon carrying zero momentum. 
These observations may be straightforwardly 
extended to the case of the three-gluon vertex  $\gh_{\alpha\mu\nu}(q,r,p)$, allowing 
for a completely analogous pictorial representation of the corresponding WI. 
In fact, the block-wise realization of the STI leads 
to a corresponding pattern for the WIs that emerges from it: 
the derivative acting on {\it any} of the blocks of $\widehat{\Pi}_{\mu\nu}(q)$
is identical to the diagrams 
comprising the associated block of $\gh_{\alpha\mu\nu}(q,r,p)$, when the corresponding momentum is set to zero. 
To the best of our knowledge, the notions described above appear for the first time 
in the literature. 

The article is organized as follows. 
 In Sec.~\ref{sec_the} we review certain pivotal properties of the BFM, and explain 
 the notion of the block-wise transversality at the level of the SDE that governs    
the background gluon propagator. 
 Sec.~\ref{sec_block} contains the main result of this work, namely
the demonstration of the block-wise realization of the STI for the case of the background three-gluon vertex.  
Then, in Sec.~\ref{sec_soft} we focus on the WIs of the BFM, their graphical representation 
in terms of zero-momentum gluon insertions, and demonstrate the block-wise realization of the 
three-gluon WI, for the operationally simplest subset of graphs.   
In Sec.~\ref{sec_disc} we summarize our findings and discuss future directions. Finally, in four Appendices we present complementary material that facilitates the perusal of the article.

%%%%%%%%%%%%%%%%%%%%%%%%%%%%%%%%%%%%%%%%%%%%%%%%

\section{General theoretical framework} 
\label{sec_the}

In this section we highlight some of the significant features of the BFM formalism that are relevant for the demonstrations that follow; 
for further details the reader is referred to the extensive literature 
on the subject, see, \eg,~\cite{Abbott:1980hw,Abbott:1983zw,Binosi:2009qm}.

({\it i})
The initial decomposition of the gauge field into $B_{\mu}^{a}$ and $Q_{\mu}^{a}$ components 
increases considerably the number of Green's functions that can be defined, which may be 
classified into three broad subsets: those 
with $B_{\mu}^{a}$ fields only, those with $Q_{\mu}^{a}$ fields only (corresponding to 
the standard Green's functions of the $R_\xi$ gauges), and mixed ones, with both $B_{\mu}^{a}$ and $Q_{\mu}^{a}$ fields. We will occasionally denote Green's functions according to the type of incoming fields, such as  
``BB'' for the case of the propagator connecting two background fields, 
or ``BBB'' for the case of the three-gluon vertex connecting three such fields.  

({\it ii}) 
The gluon propagator QQ that enters in the quantum loops will be denoted 
by  \mbox{$\Delta_{\mu\nu}^{ab}(q)=-i\delta^{ab}\Delta_{\mu\nu}(q)$}, with 
\begin{align}
    \Delta_{\mu\nu}(q)= P_{\mu\nu}(q) \Delta(q) + \xi_{\s Q}\frac{q_\mu q_\nu}{q^4}\,, \qquad P_{\mu\nu}(q) = g_{\mu\nu} - \frac{q_\mu q_\nu}{q^2} \,,
\label{QQprop}
\end{align}
whose inverse is  
\begin{align}
  \Delta_{\mu}^{\nu}(q) \Delta_{\nu\rho}^{-1}(q)=g_{\mu\rho}\,; \quad\quad     \Delta_{\nu\rho}^{-1}(q)=  \Delta^{-1}(q)  P_{\nu\rho}(q) +  \xi_{\s Q}^{-1} q_\nu q_\rho \,.
\label{invBQprop}
\end{align}
The scalar function  $\Delta(q)$ is related to the gluon self-energy \mbox{$\Pi_{\mu\nu}(q)=P_{\mu\nu}(q)\Pi(q)$} through 
\mbox{$\Delta^{-1}(q)= q^2 +i\Pi(q)$}, and $\xi_{\s Q}$ is the quantum gauge-fixing parameter. Note that 
$\xi_{\s Q}$ enters also in the tree level expressions of the vertices BQQ and BBQQ, given in Table~\ref{fig_feyback}.

({\it iii})
In what follows we will use extensively a number of three- and four-particle vertices, which 
we list here.
In particular, the relevant three-particle vertices are 
\begin{align}
    & \Gamma_{\bar{c}^m c^nQ_{\mu}^a}(r,p,q) = -gf^{mna}\Gamma_{\mu}(r,p,q)\,, \qquad 
    &\Gamma_{\!Q_{\alpha}^a Q_{\mu}^b Q_{\nu}^c}(q,r,p)  = gf^{abc}\gt_{\alpha \mu \nu}(q,r,p)\,, \nonumber \\
    & \Gamma_{\bar{c}^m c^nB_{\mu}^a}(r,p,q) =-g f^{mna}\gt_\mu(r,p,q)\,, \qquad 
    & \Gamma_{\!B_{\alpha}^a Q_{\mu}^b Q_{\nu}^c}(q,r,p)  = gf^{abc}\gt_{\alpha \mu \nu}(q,r,p)\,,
\label{3vert}
\end{align}
while the four-particle  vertices are 
\begin{align} 
    &\hspace{-0.3cm} \Gamma_{B_{\mu}^a B_{\nu}^b \bar{c}^m c^n}(q,r,p,t) = -ig^2 \gh_{\mu\nu}^{abmn}(q,r,p,t), 
    &\Gamma_{\!B_{\alpha}^a Q_{\beta}^b Q_{\mu}^c Q_{\nu}^d}(q,r,p,t)  = -ig^2\gt_{\alpha \beta \mu \nu}^{abcd}(q,r,p,t), \nonumber  \\
    & \hspace{-0.3cm}\Gamma_{B_{\mu}^a Q_{\nu}^b \bar{c}^m c^n}(q,r,p,t) = -ig^2 \gt_{\mu\nu}^{abmn}(q,r,p,t), 
    & \Gamma_{\!B_{\alpha}^a B_{\beta}^b Q_{\mu}^b Q_{\nu}^c}(q,r,p,t)  = -ig^2\gh_{\alpha \beta \mu \nu}^{abcd}(q,r,p,t), 
\label{4vert}
\end{align}
where $g$ denotes the gauge coupling constant, and $f^{abc}$ are the SU(3) structure constants.

%%%%%%%%%%%%%%%%%%%%%%%%%%%%%%%%%%%%%%%%%%%%%%%%%%%%%%%%%%%%%%%%%%%%
({\it iv}) A central quantity in our analysis is the background 
self-energy, $\widehat{\Pi}_{\mu\nu}(q)$,  related to the 
inverse background gluon propagator $\widehat{\Delta}^{-1}_{\mu\nu}(q)$ by 
\footnote{The definition of the BB propagator $\widehat{\Delta}_{\mu\nu}(q)$ requires the addition to the action of a supplementary   
gauge-fixing term, which introduces the ``classical" 
gauge-fixing parameter, $\xi_{\s C}$~\cite{Abbott:1980hw,Abbott:1983zw,Binosi:2009qm}. Note that 
this step is necessary only when
connecting the background gluon to external states in order to construct S-matrix elements, and will be omitted here.}
\be
\widehat{\Delta}^{-1}_{\mu\nu}(q)= q^2 P_{\mu\nu}(q) + i \widehat{\Pi}_{\mu\nu}(q)\,. 
\label{invself}
\ee
The gauge symmetry enforces the fundamental STI  
\be
q^{\mu} \widehat{\Pi}_{\mu\nu}(q)= 0 \,,
\label{BBtrans}
\ee
from which follows that $\widehat{\Pi}_{\mu\nu}(q) = P_{\mu\nu}(q) \widehat{\Pi}(q)$, 
where $\widehat{\Pi}(q)$ is a scalar function. Thus, \1eq{invself} may be cast in the form 
\be
\widehat{\Delta}^{-1}_{\mu\nu}(q) = P_{\mu\nu}(q) \left[q^2 + i \widehat{\Pi}(q)\right]\,. 
\label{invself_scalar}
\ee

\begin{figure}[t]
    \centering
    \includegraphics[scale=0.3]{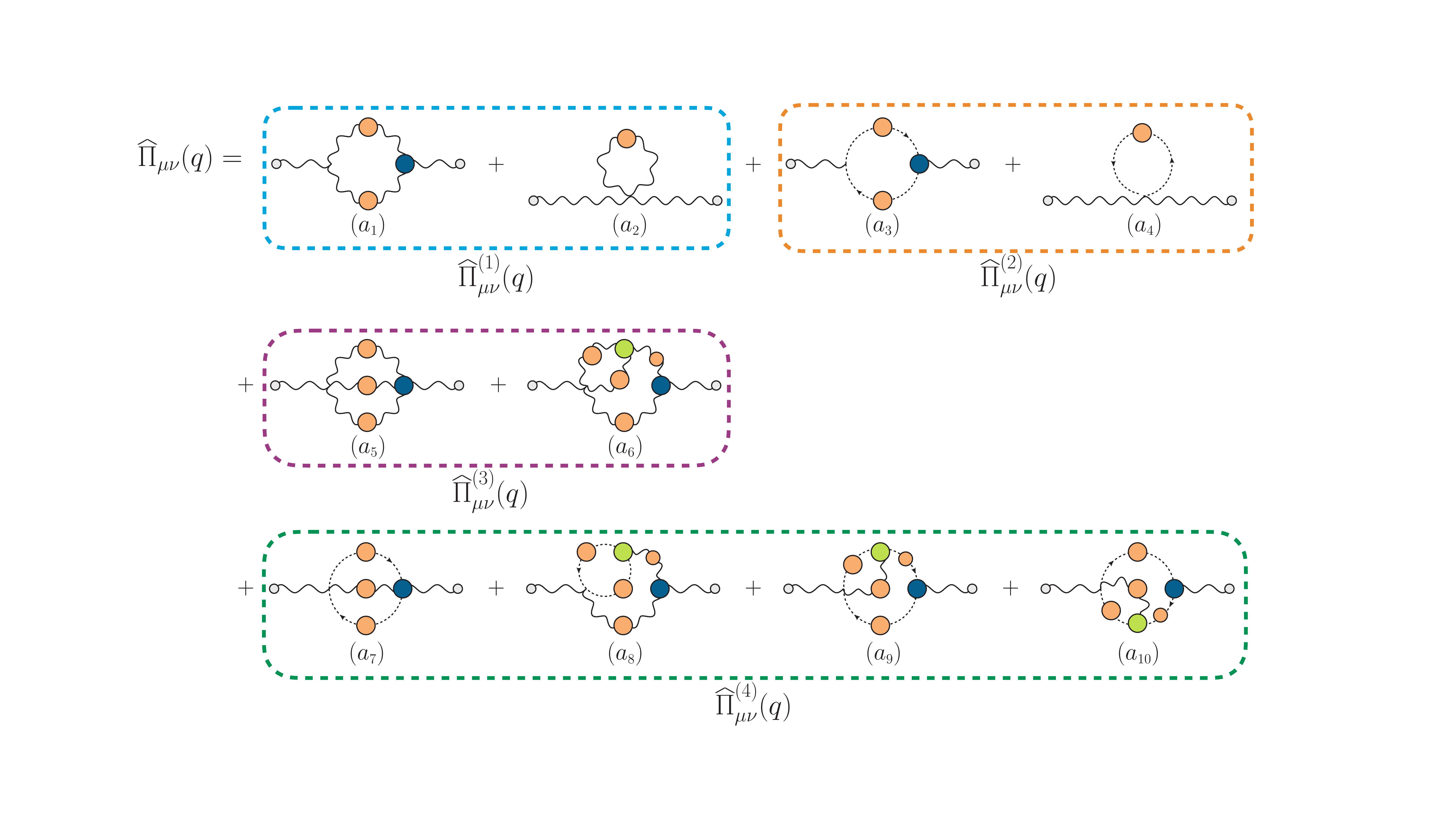}
    \caption{Diagrammatic representation of the self-energy  $\widehat\Pi_{\mu\nu}(q)$.
    The small gray circles at the end of gluon legs indicate a background field.     
    The orange and green circles represent conventional 
    fully dressed propagators  and vertices, respectively, 
    while the blue circles represent fully dressed vertices with one background gluon.}
    \label{fig_bbsde}
\end{figure}

The SDE that defines 
$\widehat{\Pi}_{\mu\nu}(q)$ is 
diagrammatically represented in \fig{fig_bbsde}. Note the separation of the 
{\it dressed} Feynman 
diagrams into the following four distinct groups: 

(1) One-loop gluonic graphs, 
enclosed in the blue box; their 
total contribution is  denoted by  $\widehat{\Pi}^{(1)}_{\mu\nu}(q)$.

(2) One-loop ghost graphs,  
enclosed in the orange box; their 
total contribution is  denoted by  $\widehat{\Pi}^{(2)}_{\mu\nu}(q)$.

(3) Two-loop gluonic graphs, enclosed in the purple box; their 
total contribution is  denoted by  $\widehat{\Pi}^{(3)}_{\mu\nu}(q)$.

(4) Two-loop ghost graphs, enclosed in the green box; their 
total contribution is  denoted by  $\widehat{\Pi}^{(4)}_{\mu\nu}(q)$.

%%%%%%%%%%%%%%%%%%%%%%%%%%%%%%%%%%%%%%%%%%%%%%%%%%%%%%%%%
%. Fig.1 -  The BB Gluon self-energy 
%%%%%%%%%%%%%%%%%%%%%%%%%%%%%%%%%%%%%%%%%%%%%%%%%%%%%%%%%
%%%%%%%%%%%%%%%%%%%%%%%%%%%%%%%%%%%%%%%%%%%%%%%%

One of the most exceptional properties of $\widehat{\Pi}_{\mu\nu}(q)$ is its block-wise transversality~\mbox{\cite{Aguilar:2006gr,Binosi:2007pi,Binosi:2008qk}}. Specifically, 
the fundamental relation given in \1eq{BBtrans}
is realized in a very special way: 
each of the four subsets of diagrams in \fig{fig_bbsde} is individually transverse,  \ie 
\be
q^{\mu} \widehat{\Pi}^{(i)}_{\mu\nu}(q)= 0\,, \qquad i=1,2,3,4 \,.
\label{blockwise}
\ee
This particular result is a direct consequence of the Abelian STIs satisfied by the fully dressed vertices entering in the diagrams comprising the 
$\widehat{\Pi}^{(i)}_{\mu\nu}(q)$~\mbox{\cite{Aguilar:2006gr,Binosi:2007pi,Binosi:2008qk}}, namely \bqq, \bcc, \bqqq\ and \bqcc, reported in Table~\ref{fig_wti} of the Appendix~\ref{apB}.

({\it v})  In order to elucidate with a simple example how 
this special transversality is enforced at the diagrammatic level, 
we consider the case of  $\ph^{(2)}_{\mu\nu}(q)$, whose diagrams 
are enclosed by the orange box of \fig{fig_bbsde}. 

The diagrams $(a_{\s{3}})$ and $(a_{\s{4}})$ are given by
\begin{align}
    (a_{\s{3}})_{\mu\nu}(q) &= \lambda  \int_k (2k-q)_\mu D(k-q)D(k)\gt_\nu(q-k,k,-q) \,,  \label{a3}\\
    (a_{\s{4}})_{\mu\nu}(q) &= 2\lambda\, g_{\mu\nu} \int_k  D(k) \,, \label{a4}
%\label{a3a4}
\end{align}
where we have used the Feynman rules given in \eq{BCC0} and \eq{BBcc0} of the Appendix~\ref{sec:App_feynman}, and factored out the trivial color structure $\delta^{ab}$ from both expressions. In addition, we have defined 
\begin{equation}
    \lambda := g^2C_{\rm A} \,,
\label{lambda}
\end{equation}
where $C_{\rm A}$ is the Casimir eigenvalue of the adjoint representation [$N$ for SU($N$)]. Furthermore, we have introduced 
\be
\int_{k} :=\frac{1}{(2\pi)^{4}}\!\int_{-\infty}^{+\infty}\!\!\mathrm{d}^4 k\,,
\label{dqd}
\ee
where the use of a symmetry-preserving regularization
scheme is implicitly assumed.

We next contract graph $(a_{\s{3}})_{\mu\nu}(q)$ by $q^\nu$, thus triggering 
the STI satisfied by \mbox{$\gt_\nu(q-k,k,-q)$}, given in \eq{BCCw}, to obtain 
\bea
    q^\nu(a_{\s{3}})_{\mu\nu}(q) &=&  \lambda \int_k (2k-q)_\mu D(k-q)D(k) \left[D^{-1}(k-q) - D^{-1}(k)\right]
 \nonumber\\   
    &=& \lambda \int_k (2k-q)_\mu \left[D(k) - D(k-q)\right]
    \label{qa3}
\nonumber\\ 
&=& -2\lambda \,q_\mu \int_k D(k)\,,
\eea
which is exactly the negative of the contraction $q^\nu(a_{\s{4}})_{\mu\nu}(q)$. Hence, 
\begin{align}
    q^\nu\left[ (a_{\s{3}})_{\mu\nu}(q) + (a_{\s{4}})_{\mu\nu}(q) \right] = 
    q^\nu \widehat{\Pi}^{(2)}_{\mu\nu}(q)=
    0\,.
\label{qa3a4}
\end{align}

We emphasize that the above strategy 
of contracting directly individual diagrams and triggering 
the corresponding STIs will be followed unaltered in the more complicated case of the three-gluon vertex 
treated in the next section. 
Note finally that 
the entire demonstration leading to \1eq{blockwise} is 
carried out for a general value of the gauge-fixing parameter $\xi_{\s Q}$~\mbox{\cite{Aguilar:2006gr,Binosi:2007pi,Binosi:2008qk}}.

\section{Block-wise STI of the three-gluon vertex } 
\label{sec_block} 

In this section we demonstrate the 
block-wise realization of the STI satisfied by the 
BBB three-gluon vertex.

\subsection{General considerations} \label{sec_gencon}

({\it i})  The one-particle irreducible three-gluon vertex, 
$\gb_{\alpha\mu\nu}^{abc}(q,r,p)$, is defined 
from the vacuum expectation value of the time ordered product of 
three background gluons (in momentum space), as 
\be
\langle 0 \vert \,T \!\left [{B}^a_{\alpha'}(q) \, {B}^b_{\mu'}(r) \,{B}^c_{\nu'} (p)\right]\!\vert 0 \rangle = g \, \gb_{\alpha\mu\nu}^{abc}(q,r,p)
\widehat{\Delta}^{\alpha}_{\alpha'}(q)\widehat{\Delta}^{\mu}_{\mu'}(r)\widehat{\Delta}^{\nu}_{\nu'}(p) \,.
\label{vev}
\ee

The three-gluon vertex $\gb_{\alpha\mu\nu}^{abc}(q,r,p)$ is naturally cast in the form  
\be
\gb^{abc}_{\alpha\mu\nu}(q,r,p) = \gh^{(0)abc}_{\alpha\mu\nu}(q,r,p) + 
\gh^{abc}_{\alpha\mu\nu}(q,r,p) \,,
\label{Bqc}
\ee
where  the tree level component $\gh^{(0)abc}_{\alpha\mu\nu}(q,r,p)=f^{abc}\gh^{(0)}_{\alpha\mu\nu}(q,r,p)$  
coincides with that of the conventional three-gluon vertex (QQQ), \ie
\be
\gh^{(0)}_{\alpha\mu\nu}(q,r,p) = \Gamma^{(0)}_{\alpha\mu\nu}(q,r,p) =  (q-r)_{\nu}g_{\alpha\mu} + (r-p)_{\alpha} g_{\mu\nu} + (p-q)_{\mu}g_{\alpha\nu}\,,
\label{BBB0}
\ee
while $\widehat{\Gamma}^{abc}_{\alpha\mu\nu}(q,r,p)$ captures all quantum corrections,
both perturbative and nonperturbative.

%%%%%%%%%%%%%%%%%%%%%%%%%%%%%%%%%%%%%%%%%%%%%%%%%%%%%%%%%
% Fig. 2 -  Block-wise structure for the BBB three-gluon vertex 
%%%%%%%%%%%%%%%%%%%%%%%%%%%%%%%%%%%%%%%%%%%%%%%%%%%%%%%%%
\begin{figure}[ht]
\centering
\includegraphics[scale=0.28]{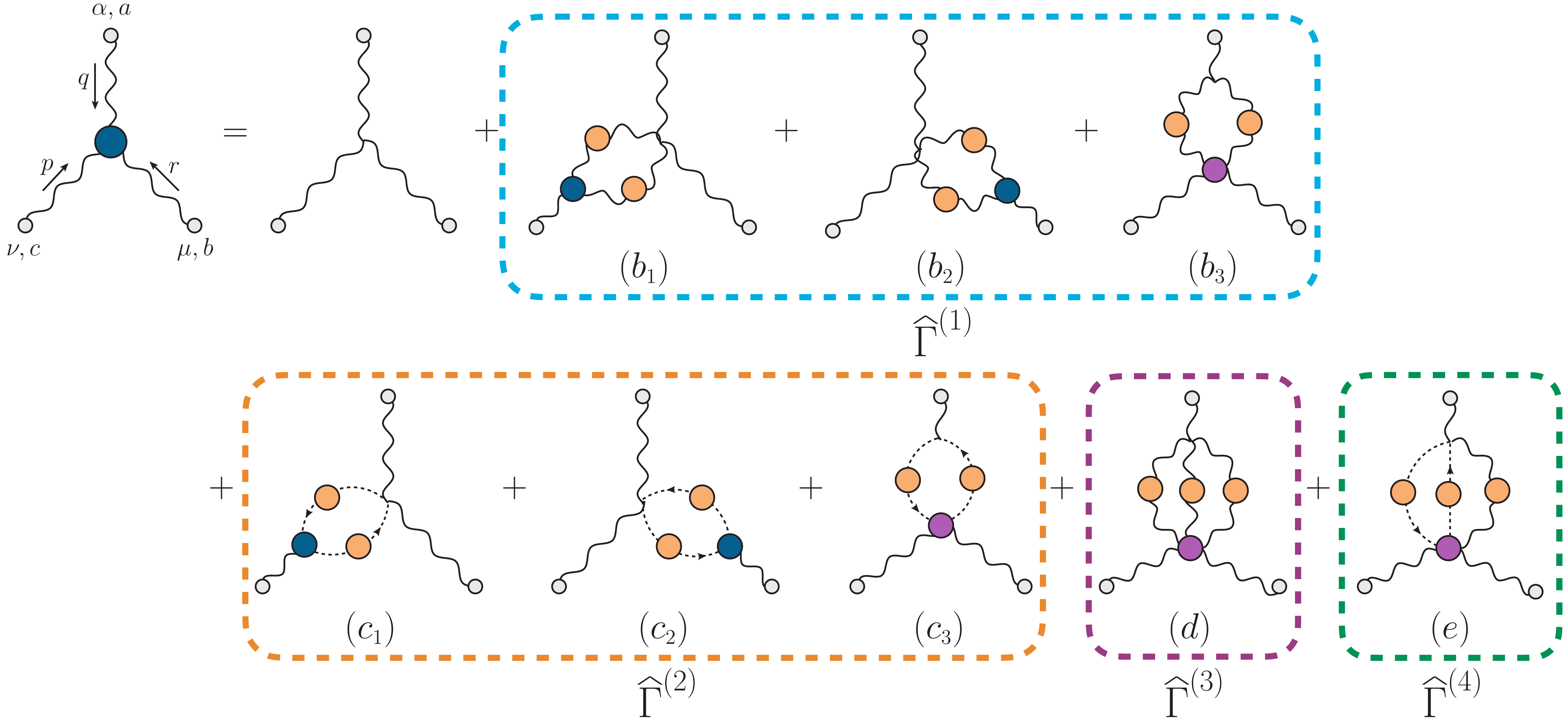}
\caption{The block-wise structure of the SDE which describes the three-gluon vertex $\gb^{abc}_{\alpha\mu\nu}(q,r,p)$ (\bbb). The blue circles represent full one-particle irreducible vertices, while the purple ones the four- and five-point scattering kernels.}
\label{fig_bbbsde}
\end{figure}
%%%%%%%%%%%%%%%%%%%%%%%%%%%%%%%%%%%%%%%%%%%%%%%%%%%%%%%%%

({\it ii}) The SDE that defines $\gb^{abc}_{\alpha\mu\nu}(q,r,p)$
is shown diagrammatically in \fig{fig_bbbsde}, written with respect to the 
gluon that carries momentum $q$; therefore, the corresponding vertices to which this leg is attached are kept at tree level.  The 
corresponding Feynman diagrams have been classified in four blocks, 
applying the exact same criterion as in 
the case 
of $\widehat{\Pi}_{\mu\nu}(q)$, and employing 
the same color code for the individual boxes as in  
\fig{fig_bbsde}. Thus, 
\be
\widehat{\Gamma}^{abc}_{\alpha\mu\nu}(q,r,p) = \sum_{i=1}^4 
\gh^{(i)abc}_{\alpha\mu\nu}(q,r,p)\,, 
\ee
where, as shown in  \fig{fig_bbbsde}, the four blocks are comprised by the diagrams (suppressing indices)  
\be
\gh^{(1)}= (b_{\s{1}})+(b_{\s{2}})+(b_{\s{3}}) \,,  \qquad
\gh^{(2)}= (c_{\s{1}})+(c_{\s{2}})+(c_{\s{3}})\,, \qquad\
\gh^{(3)}=  (d) \,, \qquad
\gh^{(4)}= (e) \,.
\label{3gblocks}
\ee

({\it iii}) It is well-known that $\gb^{abc}_{\alpha\mu\nu}(q,r,p)$ 
satisfies the Abelian STI~\cite{Cornwall:1989gv,Aguilar:2006gr,Binosi:2009qm}
\be
\label{eq_WTIBBB}
p^\nu \gb^{abc}_{\alpha\mu\nu}(q,r,p) = f^{cae}\left[\widehat{\Delta}^{be}_{\alpha\mu}(r)\right]^{-1} - f^{bce}\left[\widehat{\Delta}^{ae}_{\alpha\mu}(q)\right]^{-1}\,,
\ee
and cyclic permutations thereof. 
The STI of \1eq{eq_WTIBBB} 
may be obtained by means of 
formal manipulations of the BFM  
generating functional, or 
simply from 
the STI of the 
conventional QQQ vertex~\cite{Marciano:1977su,Ball:1980ax,Davydychev:1996pb},
by setting all ghost-related 
contributions to their tree level values. 

From \1eq{BBB0} it is elementary to show that 
\begin{align}
    p^{\nu} \gh^{(0)}_{\alpha\mu\nu}(q,r,p) = r^2 P_{\alpha\mu}(r) -q^2 P_{\alpha\mu}(q)\,.   
\end{align}
Then, from \3eqs{invself}{Bqc}{eq_WTIBBB} follows that 
\be
p^{\nu} \widehat{\Gamma}^{abc}_{\alpha\mu\nu}(q,r,p) = if^{cae}\widehat{\Pi}^{be}_{\alpha\mu}(r) -if^{bce}\widehat{\Pi}^{ae}_{\alpha\mu}(q) \,.
\label{stiqc}
\ee
({\it iv}) The central observation of the 
present study is that, as happens in the case of 
\1eq{BBtrans}, the STI in \1eq{stiqc}
admits a block-wise realization. Specifically,  
as we will demonstrate in this section, 
\be
p^{\nu} \gh^{(i)abc}_{\alpha\mu\nu}(q,r,p) = if^{cae}\widehat{\Pi}^{(i)be}_{\alpha\mu}(r) -if^{bce}\widehat{\Pi}^{(i)ae}_{\alpha\mu}(q)\,, \qquad i=1,2,3,4 \,.
\label{stiblocks}
\ee

In diagrammatic terms, \1eq{stiblocks} states 
that the contraction by $p^{\nu}$ of the diagrams within  
a given block (color) in \fig{fig_bbbsde} 
generates the difference between diagrams within  
the corresponding block (color) of \fig{fig_bbsde}.
In fact, the validity of \1eq{stiblocks} will be demonstrated by 
acting with $p^{\nu}$ on vertex diagrams,   
and exploiting the STIs triggered by this contraction in order to cast the result 
in the form of self-energy contributions.

({\it v}) Note that 
the diagrams shown in \fig{fig_bbbsde} have a factor $g$ removed from them, which cancels 
against the $g$ appearing in 
the definition of the three-gluon vertex in \eq{vev}.
%the transition from $\gh_{\alpha\mu\nu}^{abc}$ to $\gh_{\alpha\mu\nu}(q,r,p)$, according to \1eq{Bfull}. 
In addition a factor of $i$ will 
be factored out, which will cancel against the explicit $i$ appearing on the r.h.s. of \1eq{stiblocks}. 
Thus, on the r.h.s. of all intermediate formulas  
will appear directly the self-energy diagrams $(a_i)$ of Fig.~\ref{fig_bbsde}. 
Furthermore, with the exception of Sec.~\ref{sec_soft}, we will suppress the 
argument $(q,r,p)$ in all vertex graphs.

({\it vi})
We introduce the definitions 
\begin{align}
    h_1^{abmn} = f^{abe}f^{mne} \,, \qquad h_2^{abcde} = f^{abm}f^{cmn}f^{dne} \,.
\end{align}
Note that \mbox{$h_1^{abmn} = h_1^{banm} $} and \mbox{$h_2^{abcde}=h_2^{baced}$}.

\medskip

({\it vii})
We introduce  the short-hand notation 
\begin{align}   
\label{eq_Rs}
    & \yy^{\alpha\beta}_\mu(r,p) \equiv \Delta^{\alpha\sigma}(r)\Delta^{\beta \rho}(p)\Gamma_{\mu\sigma\rho}(-r-p,r,p), \, &  \yy_\mu(r,p) \equiv D(r)D(p)\Gamma_\mu(r,p,-r-p), \nonumber \\ 
    & \yt^{\alpha\beta}_\mu(r,p) \equiv \Delta^{\alpha\sigma}(r)\Delta^{\beta \rho}(p)\gt_{\mu\sigma\rho}(-r-p,r,p), \, & \yt_\mu(r,p) \equiv D(r)D(p)\gt_\mu(r,p,-r-p).
\end{align}
The special relations 
\begin{align}   \label{eq_prop1}
    \int_k \yt^{\sigma\beta}_\mu (k,r-k) = \int_k \yt_\mu (k,r-k) =0 \,,
\end{align}
will be employed in the analysis that follows. Their validity may be 
established by appealing to the Bose symmetry of the \bqq\ vertex with respect to its two quantum legs or the 
ghost-antighost symmetry of the \bcc\ vertex, and the change 
of integration variable $r-k \to k$.
An alternative demonstration proceeds by noting that  
\be
\int_k \yt_\mu (k,r-k) =  \frac{r_\mu}{r^2} I(r^2) \,,
\ee 
with 
\begin{align}
    I(r^2) = & \int_k D(k)D(r-k) \, r^\mu \gt_\mu(k,r-k,-r)  \nonumber \\
    =& \int_k D(k)D(r-k)\left[ D^{-1}(k)-D^{-1}(r-k) \right] \nonumber \\
    =& \int_k \left[D(r-k) - D(k) \right] =0\,, 
\end{align}
where the STI of \1eq{BCCw} was used. The first 
relation in \1eq{eq_prop1} may be proved in the exact same way, employing the STI of \1eq{BQQw}.

({\it viii}) Lastly, from now on we adopt the convention that 1PI vertices containing at least one background gluon will be represented diagrammatically by a blue circle (see, \eg,  the BBQQ vertex in diagram $(b_{\s{3,2}})$ of \fig{fig_1gl}).

\subsection{One-loop gluonic sector ({\it first block})} \label{sec_1gl} 

We begin by 
considering the one-loop gluonic vertex graphs,  
namely the set $\{(b_{\s{1}}), (b_{\s{2}}), (b_{\s{3}})\}$, 
enclosed in the blue box of \fig{fig_bbbsde}.

As a first step, we recognize that,  by virtue of \1eq{eq_prop1}, diagrams $(b_{\s{1}})$ and $(b_{\s{2}})$ vanish,
\begin{align}
    (b_{\s{1}})^{abc}_{\alpha\mu\nu} &= \frac{g^2}{2} f^{ced} \!\! \int_k \! \gh_{\alpha\mu\beta\sigma}^{(0)abde} \yt^{\sigma\beta}_\nu (-k,k-p) =0\,, \nonumber \\
    (b_{\s{2}})^{abc}_{\alpha\mu\nu} &= \frac{g^2}{2} f^{bed}\!\! \int_k \! \gh_{\alpha\nu\beta\sigma}^{(0)acde} \yt^{\sigma\beta}_\mu (-k,k-r) =0\,, 
\label{bnull}
\end{align}
since $\gh_{\alpha\mu\beta\sigma}^{(0)abde}$ is momentum-independent,  
and may be pulled out of the integral sign.

%%%%%%%%%%%%%%%%%%%%%%%%%%%%%%%%%%%%%%%%%%%%%%%%%%%%%%%%%%%%%%%%%%%%%
%. Fig.3 - Skeleton expansion of diagram  b3 -  Scattering kernel 4 gluons 
%%%%%%%%%%%%%%%%%%%%%%%%%%%%%%%%%%%%%%%%%%%%%%%%%%%%%%%%%%%%%%%%%%%%%
\begin{figure}[t]
    \centering
    \includegraphics[scale=0.52]{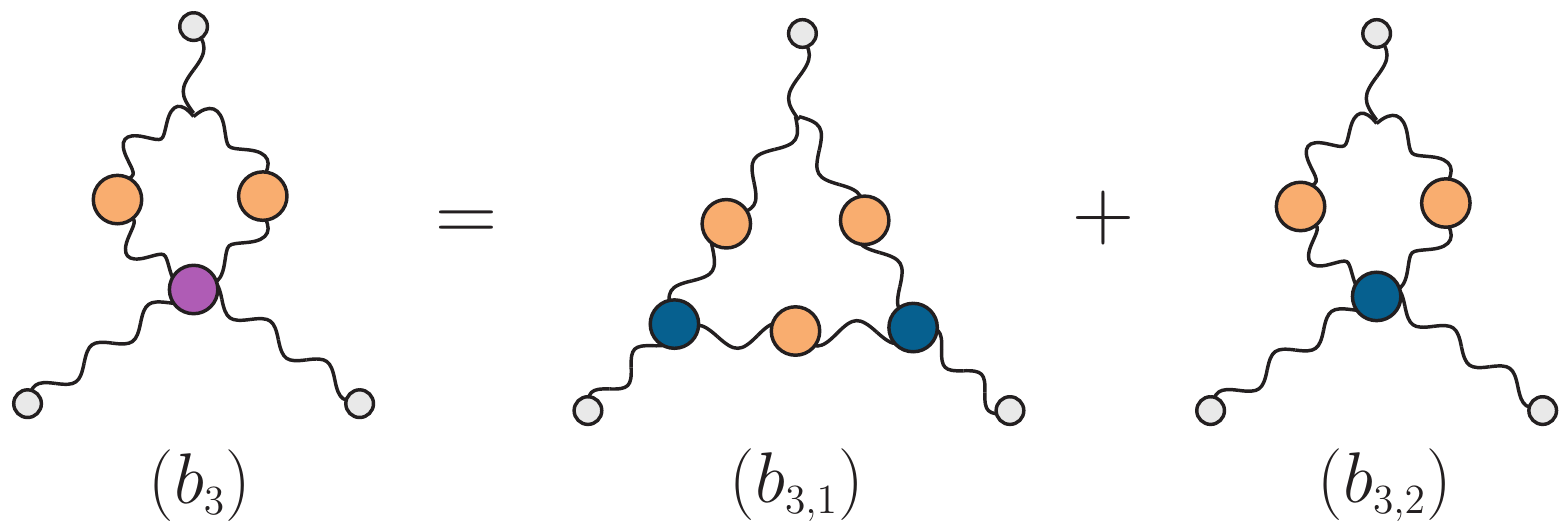}
    \caption{ The two contributions which arise from diagram $(b_{\s{3}})$ of \fig{fig_bbbsde} after performing the skeleton expansion of the four-gluon scattering kernel (purple blob).}
    \label{fig_1gl}
\end{figure}
%%%%%%%%%%%%%%%%%%%%%%%%%%%%%%%%%%%%%%%%%%%%%%%%%%%%%%%%%

Diagram $(b_{\s{3}})$ has two contributions, 
\begin{align}
    (b_{\s{3}})^{abc}_{\alpha\mu\nu} = (b_{\s{3,1}})^{abc}_{\alpha\mu\nu} + (b_{\s{3,2}})^{abc}_{\alpha\mu\nu} \,,
\end{align}
given in \fig{fig_1gl}, with
\begin{align}
    (b_{\s{3,1}})^{abc}_{\alpha\mu\nu} &= -\frac{\lambda}{2} f^{abc} \!\!\int_k \! \gt_{\alpha\beta\sigma}^{(0)}(q,k-q,-k) \Delta^{\rho\sigma}(k)\yt_{\mu}^{\lambda\beta}(k+p,q-k) \gt_{\nu\rho\lambda}(p,k,-k-p)  \,, \nonumber \\
    (b_{\s{3,2}})^{abc}_{\alpha\mu\nu} &= \frac{g^2}{2} f^{ade}\!\! \int_k \! \gt_{\alpha\beta\sigma}^{(0)}(q,k-q,-k) \Delta^{\beta\rho}(k-q) \Delta^{\sigma\lambda}(k)\gh^{cbde}_{\nu\mu\rho\lambda}(p,r,q-k,k)  \,.
\end{align}

The contraction $p^\nu(b_{\s{3}})^{abc}_{\alpha\mu\nu}$ may be 
evaluated using the STIs in 
\2eqs{BQQw}{BBQQw} and a moderate amount of algebra, yielding 
\begin{align} \label{eq_d3final}
    p^\nu(b_{\s{3}})^{abc}_{\alpha\mu\nu} = &-\frac{\lambda}{2}f^{abc} \!\!\int_k \! \gt_{\alpha\beta\sigma}^{(0)}(q,k-q,-k) \left[\yt^{\sigma\beta}_{\mu}(k,q-k) +
    \yt^{\sigma\beta}_{\mu}(-k-p,k-q) \right] \,.
\end{align}
The first term in \1eq{eq_d3final}
is exactly  $-f^{bce}(a_{\s{1}})^{ae}_{\alpha\mu}(q)$. 
The second term, after $k\rightarrow-k-p$, generates  $f^{cae}(a_{\s{1}})^{be}_{\alpha\mu}(r)$; note, in particular, that 
\begin{align}
    \gt_{\alpha\beta\sigma}^{(0)}(q,k-q,-k) \rightarrow -\gt_{\alpha\beta\sigma}^{(0)}(r,k-r,-k) + 2p_\beta g_{\alpha\sigma}-p_\alpha g_{\beta\sigma} - (\xi_{\s Q}^{-1}+1)p_\sigma g_{\alpha\beta} \,,
\end{align}
where, due to \1eq{eq_prop1}, the last three terms give vanishing contributions. 
Thus, one arrives at 
\begin{align}   \label{eq_1glfinal}
    p^\nu (b_{\s{3}})^{abc}_{\alpha\mu\nu}=f^{cae}(a_{\s{1}})^{be}_{\alpha\mu}(r) - f^{bce}(a_{\s{1}})^{ae}_{\alpha\mu}(q) \,.
\end{align}
At this point, we add and subtract on the r.h.s. of \1eq{eq_1glfinal} the momentum-independent 
seagull graph $(a_{\s{2}})$, to obtain 
\begin{align}   \label{eq_1glfinal2}
    p^\nu (b_{\s{3}})^{abc}_{\alpha\mu\nu}= f^{cae}[(a_{\s{1}}) + (a_{\s{2}})]_{\alpha\mu}^{be}(r) - f^{bce}
    [(a_{\s{1}}) + (a_{\s{2}})]_{\alpha\mu}^{ae}(q) \,,
\end{align}
which, in view of \1eq{bnull},  is precisely \1eq{stiblocks} for $i=1$.

\subsection{One-loop ghost sector ({\it second block})} \label{sec_1gh}

%%%%%%%%%%%%%%%%%%%%%%%%%%%%%%%%%%%%%%%%%%%%%%%%%%%%%%%%%%%%%%%%%%%%%%%%%%%%%%%%%
% Fig. 4   Skeleton expansion of diagram  c3 -  Scattering kernel 2 gluons-2 ghosts 
%%%%%%%%%%%%%%%%%%%%%%%%%%%%%%%%%%%%%%%%%%%%%%%%%%%%%%%%%%%%%%%%%%%%%%%%%%%%%%%%
\begin{figure}[t]
    \centering
    \includegraphics[scale=0.52]{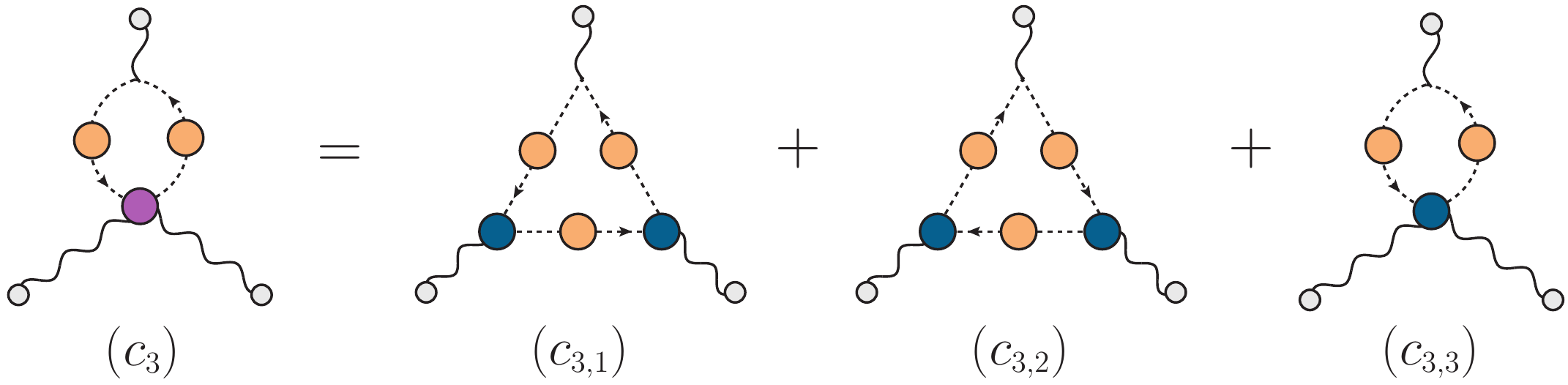}
    \caption{ The three contributions which emerge from the diagram $(c_{\s{3}})$ shown in the \fig{fig_bbbsde}, after implementing the skeleton expansion of the four-point scattering kernel (purple blob) formed by two background gluons with a ghost-antighost pair.}
    \label{fig_1gh}
\end{figure}
%%%%%%%%%%%%%%%%%%%%%%%%%%%%%%%%%%%%%%%%%%%%%%%%%%%%%%%%%%

We next focus on the one-loop ghost graphs, 
forming the set $\{(c_{\s{1}}), (c_{\s{2}}), (c_{\s{3}})\}$, 
enclosed in the orange box of \fig{fig_bbbsde}. 
The demonstration that follows is completely analogous to that of the 
previous subsection. 

Due to \1eq{eq_prop1}, 
diagrams $(c_{\s{1}})$ and $(c_{\s{2}})$ vanish,
\bea
    (c_{\s{1}})^{abc}_{\alpha\mu\nu} &=& g^2 f^{edc} \!\!\int_k\! \gh_{\alpha\mu}^{(0)abde}\yt_\nu(-k,k-p) = 0 \,, \nonumber\\
    (c_{\s{2}})^{abc}_{\alpha\mu\nu} &=&  g^2 f^{edb} \!\!\int_k\!\gh_{\alpha\nu}^{(0)acde} \yt_\mu(-k,k-r) = 0 \,,
\label{ghnull}
\eea
and we only need to consider the contraction of graph $(c_{\s{3}})$. This diagram contains 
three contributions, depicted in \fig{fig_1gh}, 
\begin{align}   \label{eq_c3}
    (c_{\s{3}})^{abc}_{\alpha\mu\nu}= \sum_{j=1}^3 (c_{\s{3,j}})^{abc}_{\alpha\mu\nu} \,,
\end{align}
with 
\begin{align}   \label{eq_cs}
    (c_{\s{3,1}})^{abc}_{\alpha\mu\nu} &= -\frac{\lambda}{2} f^{abc}\!\! \int_k \!(2k-q)_\alpha D(k-q) \yt_\mu(k,-k-r) \gt_\nu(k+r,q-k,p) \nonumber \,, \\ 
    (c_{\s{3,2}})^{abc}_{\alpha\mu\nu} &=  -\frac{\lambda}{2} f^{abc}\!\! \int_k \! (2k-q)_\alpha D(k-q) \yt_\mu(-k-r,k) \gt_\nu(q-k,k+r,p) \nonumber \,, \\ 
    (c_{\s{3,3}})^{abc}_{\alpha\mu\nu} &= -g^2 f^{eda} \!\!\int_k\! (2k-q)_\alpha  D(k) D(q-k)  \gh_{\mu\nu}^{bcde}(r,p,q-k,k) \,.
\end{align}

The contraction of diagram $(c_{\s{3}})$ by the momentum $p^\nu$ 
activates the STIs of  \2eqs{BCCw}{BBccw}, and one obtains.
\begin{align} \label{eq_d7final}
    &p^\nu(c_{\s{3}})^{abc}_{\alpha\mu\nu} =  -\lambda f^{abc} \!\!\int_k \! (2k-q)_\alpha \! \left[\yt_\mu(k-q,-p-k)+\yt_\mu(q-k,k)  \right]  \,.
\end{align}
The second term of \1eq{eq_d7final} is simply 
$-f^{bce}(a_{\s{3}})^{ae}_{\alpha\mu}(q)$, while the first term, after 
the shift $k\rightarrow-k-p$ and use of \1eq{eq_prop1}, 
furnishes  $f^{cae}(a_{\s{3}})^{be}_{\alpha\mu}(r)$. 
Thus, we conclude that 
\begin{align}   \label{eq_1gh0}
    p^\nu (c_{\s{3}})^{abc}_{\alpha\mu\nu}=f^{cae}(a_{\s{3}})^{be}_{\alpha\mu}(r) -f^{bce}(a_{\s{3}})^{ae}_{\alpha\mu}(q) \,, 
\end{align}
which, after adding and subtracting the momentum-independent ($a_4$), is tantamount to the validity of 
\1eq{stiblocks} for $i=2$.

\subsection{Two-loop gluonic sector ({\it third block})} \label{sec_2gl} 

\begin{figure}[h]
    \centering
    \includegraphics[scale=0.52]{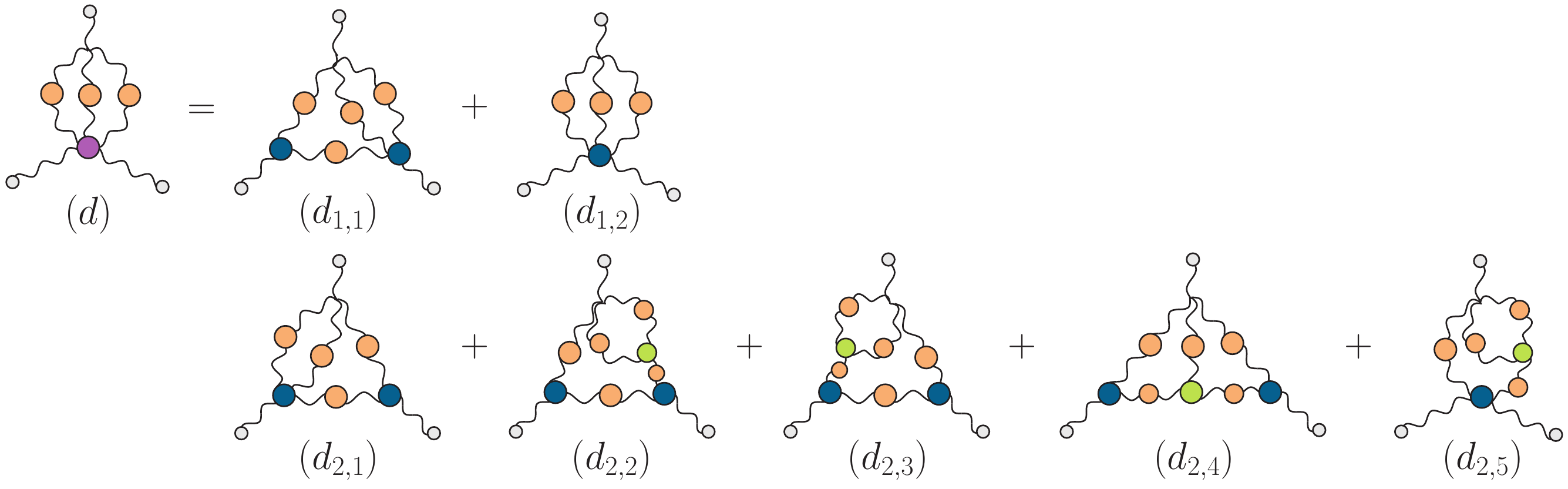}
    \caption{ The seven contributions originating from diagram $(d)$ of the \fig{fig_bbbsde}. This group splits into two subsets: the first composed by the diagrams
     $\{(d_{\s{1,1}}), (d_{\s{1,2}})\}$ and the second by $\{(d_{\s{2,1}}), (d_{\s{2,2}}), (d_{\s{2,3}}),(d_{\s{2,4}}), (d_{\s{2,5}})\}$.}
    \label{fig_2gl}
\end{figure}

We turn to the two-loop dressed gluonic contributions, 
contained within the diagram $(d)$, enclosed by the purple box in \fig{fig_bbbsde}. 
To that end, in \fig{fig_2gl} we show the individual diagrams 
that emerge upon implementing  
the skeleton expansion of the five-gluon kernel (purple circle) inside  $(d)$. 
Note that all vertices appearing in these graphs 
are one-particle irreducible. 

%%%%%%%%%%%%%%%%%%%%%%%%%%%%%%%%%%%%%%%%%%%%%%%%%%%%%%%%%
%Fig. 5  Skeleton expansion of diagram  d -   5-gluon scattering kernel 
%%%%%%%%%%%%%%%%%%%%%%%%%%%%%%%%%%%%%%%%%%%%%%%%%%%%%%%%%
%%%%%%%%%%%%%%%%%%%%%%%%%%%%%%%%%%%%%%%%%%%%%%%%%%%%%%%%

The diagrams in \fig{fig_2gl} are naturally organized into two subsets, 
\begin{align}
    (d)^{abc}_{\alpha\mu\nu} = (d_{\s{1}})^{abc}_{\alpha\mu\nu} + (d_{\s{2}})^{abc}_{\alpha\mu\nu}   \,, 
\end{align}
with
\begin{align}   \label{cdecomp}
    (d_{\s{1}})^{abc}_{\alpha\mu\nu} = \sum_{j=1}^{2} (d_{\s{1,j}})^{abc}_{\alpha\mu\nu} \,, \qquad  
    (d_{\s{2}})^{abc}_{\alpha\mu\nu} =\sum_{j=1}^{5} (d_{\s{2,j}})^{abc}_{\alpha\mu\nu} \,, 
\end{align}
where 
\begin{align} 
    (d_{\s{1,1}})^{abc}_{\alpha\mu\nu} &= \frac{ig^4}{2} f^{cen} \!\! \int_k \int_l \gt_{\alpha\beta\sigma\rho}^{(0)adme}\Delta^{\beta\tau}(k) \Delta^{\sigma\eta}(s) \yt_\nu^{\rho\lambda}(l,-p-l) \gt_{\mu\lambda\eta\tau}^{bnmd}(r,l+p,s,k)  \,, \nonumber \\
    (d_{\s{1,2}})^{abc}_{\alpha\mu\nu} &= -\frac{ig^4}{6} \!\! \int_k \int_l \gt_{\alpha\beta\sigma\rho}^{(0)adme}\Delta^{\beta\tau}(k) \Delta^{\sigma\eta}(s) \Delta^{\rho\lambda} (l) \gh_{\nu\mu\lambda\eta\tau}^{cbemd}(p,r,l,s,k) \,,
\end{align}
and
\begin{align}  
    (d_{\s{2,1}})^{abc}_{\alpha\mu\nu} &= \frac{ig^4}{2} f^{ben} \!\! \int_k \int_l \gt_{\alpha\beta\sigma\rho}^{(0)adme}\Delta^{\beta\tau}(k) \Delta^{\sigma\eta}(s) \yt_\mu^{\rho\lambda}(l,-r-l) \gt_{\nu\lambda\eta\tau}^{cnmd}(p,l+r,s,k)  \,, \nonumber \\
    (d_{\s{2,2}})^{abc}_{\alpha\mu\nu} &= \frac{ig^4}{2} h_2^{dmbec} \!\! \int_k \int_l \gt_{\alpha\beta\sigma\rho}^{(0)adme} \Delta^{\rho\lambda} (l) \yy_\tau^{\sigma\beta}(s,k) \yt_\mu^{\eta\tau}(l+p,q-l) \gt_{\nu\lambda\eta}(p,l,-l-p) \,, \nonumber \\
    (d_{\s{2,3}})^{abc}_{\alpha\mu\nu} &= \frac{ig^4}{2} h_2^{dmceb}  \!\! \int_k \int_l \gt_{\alpha\beta\sigma\rho}^{(0)adme}\Delta^{\beta\tau}(k) \yy_\tau^{\eta\sigma}(l-q,s) \yt_\mu^{\rho\lambda}(l,-r-l) \gt_{\nu\lambda\eta}(p,l+r,q-l)  \,, \nonumber \\
    (d_{\s{2,4}})^{abc}_{\alpha\mu\nu} &= ig^4h_2^{cdmbe} \!\! \int_k \int_l \gt_{\alpha\beta\sigma\rho}^{(0)adme} \Delta^{\beta\eta}(k) \yt_\mu^{\rho\tau}(l,-r-l) \yy_\tau^{\sigma\lambda}(s,k+p) \gt_{\nu\lambda\eta}(p,-k-p,k)  \,, \nonumber \\
    (d_{\s{2,5}})^{abc}_{\alpha\mu\nu} &= \frac{ig^4}{2} f^{mdn} \!\! \int_k \int_l \gt_{\alpha\beta\sigma\rho}^{(0)adme}\Delta^{\beta\tau}(k)  \Delta^{\rho\lambda} (l)   \yy_\tau^{\eta\sigma}(l-q,s) \gh^{cben}_{\nu\mu\lambda\eta}(p,r,l,q-l) \,. 
\end{align}

This particular separation is motivated by the observation that the contraction by $p^\nu$ of each subset generates a concrete term of the STI, namely 
\begin{align}   
    p^\nu (d_{\s{1}})^{abc}_{\alpha\mu\nu}=f^{cae}(a_{\s{5}})^{be}_{\alpha\mu}(r) - f^{bce}(a_{\s{5}})^{ae}_{\alpha\mu}(q) \,, \nonumber \\
    p^\nu (d_{\s{2}})^{abc}_{\alpha\mu\nu}=f^{cae}(a_{\s{6}})^{be}_{\alpha\mu}(r) - f^{bce}(a_{\s{6}})^{ae}_{\alpha\mu}(q) \,, \label{eq_2gl2}
\end{align}
where the self-energy diagrams $(a_{\s{5}})$ and $(a_{\s{6}})$
(purple box in \fig{fig_bbsde})
are given by
\begin{align}
    (a_{\s{5}})^{ab}_{\alpha\mu}(q) &= -\frac{ig^4}{6} \!\! \int_k \int_l \gt_{\alpha\beta\sigma\rho}^{(0)adme} \Delta^{\beta\tau} (k) \Delta^{\sigma\eta}(s) \Delta^{\rho\lambda}(l) \gt_{\mu\lambda\eta\tau}^{bemd}(-q,l,s,k) \nonumber \\
    &=  ig^4 h_1^{adme} \int_k \int_l \Delta^{\beta\tau} (k) \Delta^{\eta}_\beta(s) \Delta^{\lambda}_\alpha(l) \gt_{\mu\lambda\eta\tau}^{bemd}(-q,l,s,k) \,, \nonumber \\
    (a_{\s{6}})^{ab}_{\alpha\mu}(q) &= \frac{ig^4}{2}   h_1^{bedm} \!\! \int_k \int_l \gt_{\alpha\beta\sigma\rho}^{(0)adme}  \yt_{\mu}^{\rho \lambda}(l,q-l) \yy_\lambda^{\sigma\beta}(s,k) \label{eq_a6} \,,
\end{align}
with $s=q-k-l$. 
In passing from the first to 
the second expression for 
$(a_{\s{5}})$, the explicit form  
of $\gt_{\alpha\beta\sigma\rho}^{(0)adme}$, 
given in \eq{BQQQ0}, has been used.

The contraction of the momentum $p^\nu$ with the above diagrams 
will activate a series of STIs, which
will furnish the desired structures, together 
with a considerable number of terms 
that will cancel exactly among each other. In what follows, we briefly outline  
how this calculation may be best organized.

We begin with the diagrams comprising $(d_{\s{1}})^{abc}_{\alpha\mu\nu}$. 
The action of $p^\nu$ on $(d_{\s{1,1}})^{abc}_{\alpha\mu\nu}$  leads to the contraction $p^\nu \yt_\nu^{\rho\lambda}(l,-p-l)$, triggering the STI of \eq{BQQw}, and yielding
\begin{align}
    p^\nu\yt_\nu^{\rho\lambda}(l,-p-l)=\Delta^{\rho\alpha}(l)\Delta^{\lambda\beta}(l+p) \left[\Delta^{-1}_{\alpha\beta}(l+p)- \Delta^{-1}_{\alpha\beta}(l) \right] = \Delta^{\rho\lambda}(l) - \Delta^{\rho\lambda}(l+p) \,.
\end{align}
The second term in the above expression will give
\begin{align}
   \hspace{-0.2cm}  p^\nu(d_{\s{1,1}})^{abc}_{\alpha\mu\nu} &= -\frac{ig^4}{2} f^{cen} \!\! \int_k \int_l \gt_{\alpha\beta\sigma\rho}^{(0)adme}\Delta^{\beta\tau}(k) \Delta^{\sigma\eta}(s) \Delta^{\rho\lambda}(l+p) \gt_{\mu\lambda\eta\tau}^{bnmd}(r,l+p,s,k) + \cdots, 
\end{align}
where the ellipsis denotes terms that do not contribute to the r.h.s. of \eq{eq_2gl2}.  
We now change the 
integration variables as 
$k \rightarrow -k$ and $l \rightarrow -l-p$, 
and exploit Lorentz invariance to 
replace 
$\gt_{\mu\lambda\eta\tau}^{bnmd}(r,-l,-t,-k) \rightarrow 
\gt_{\mu\lambda\eta\tau}^{bnmd}(-r,l,t,k)$. 
Then, substituting the Feynman rule of \eq{BQQQ0} and using the Bose symmetry of the vertices,
we arrive at 
\begin{align}   \label{eq_d11}
    p^\nu(d_{\s{1,1}})^{abc}_{\alpha\mu\nu} &= ig^4 f^{cae} h_1^{edmn} \int_k \int_l \Delta^{\beta\tau} (k) \Delta^{\eta}_\beta(t) \Delta^{\lambda}_\alpha(l) \gt_{\mu\lambda\eta\tau}^{bnmd}(-r,l,t,k) + \cdots \,,
\end{align}
which is precisely $f^{cae}(a_{\s{5}})^{be}_{\mu\nu}(r)$, in the form given in  the second line 
of  \eq{eq_a6}.

Similarly, the contraction of $(d_{\s{1,2}})^{abc}_{\alpha\mu\nu}$ 
by $p^\nu$ activates the STI for the five-point function $\gh_{\nu\mu\lambda\eta\tau}^{cbemd}(p,r,l,s,k)$, given in \eq{BBQQQw},  namely
\begin{align}
    p^\nu\gh^{cbemd}_{\nu\mu\lambda\eta\tau}(p,r,l,s,k) &= f^{bcx}\gt_{\mu\lambda\eta\tau}^{xemd}(r+p,l,s,k) + f^{ecx}\gt_{\mu\lambda\eta\tau}^{bxmd}(r,l+p,s,k) \nonumber\\
    &+f^{mcx}\gt_{\mu\lambda\eta\tau}^{bexd}(r,p,s+p,k) + f^{dcx}\gt_{\mu\lambda\eta\tau}^{bemx}(r,p,s,k+p) \,. 
\label{pgh}
\end{align}
Note that only the first term on the r.h.s. of \eq{pgh} 
contains the four-point function with a $q$ entry, since $r+p=-q$. Thus, one gets  
\begin{align} \label{eq_d12}
    p^\nu (d_{\s{1,2}})^{abc}_{\alpha\mu\nu} = \frac{ig^4}{6} f^{bcx} \int_k \int_l \gt_{\alpha\beta\sigma\rho}^{(0)adme}&\Delta^{\beta\tau}(k) \Delta^{\sigma\eta}(s)  \Delta^{\rho\lambda} (l) \gt_{\mu\lambda\eta\tau}^{xemd}(-q,l,s,k) + \cdots \,,
\end{align}
which is precisely $f^{bce}(a_{\s{5}})^{ae}(q)$. 

After appropriate changes in the integration variables and 
judicious use of Bose symmetry, one may show that 
all remaining terms, denoted by the ellipses in \2eqs{eq_d11}{eq_d12}, cancel against each other.
Thus, one is left with 
the first equation in \eq{eq_2gl2}.

A similar line of reasoning reveals that the term  $f^{cae}(a_{\s{6}})^{be}_{\alpha\mu}(r)$ originates from the contraction of $p^\nu$ with the diagrams $(d_{\s{2,2}})$ and $(d_{\s{2,4}})$ of the second group. 
Specifically, one triggers the STI of \eq{BQQw} to obtain
\begin{align}   
\label{eq_d22_d24}
 p^\nu \left[(d_{\s{2,2}})\!+\!(d_{\s{2,4}})\right]^{abc}_{\alpha\mu\nu}\!= \frac{ig^4}{2}\left[ h_2^{dmbec} \!+\! 2h_2^{cdmbe}\right] \!\! \int_k \!\int_l  \gt_{\alpha\beta\sigma\rho}^{(0)adme}\yt_\mu^{\rho\lambda}(l,r-l)  \yy_\lambda^{\sigma\beta}(t,k) + \cdots\,.
\end{align}
Using the Feynman rule given by Eq.~\eqref{BQQQ0} for the tree level vertex one gets that
\begin{align}
    \frac{ig^4}{2}\left[ h_2^{dmbec} \!+\! 2h_2^{cdmbe}\right]  \gt_{\alpha\beta\sigma\rho}^{(0)adme} = \frac{ig^4}{2}   f^{cax} h_1^{bedm}  \gt_{\alpha\beta\sigma\rho}^{(0)xdme} + \cdots \,;
\end{align}
the substitution of the 
first term into \eq{eq_d22_d24} gives precisely $f^{cae}(a_{\s{6}})^{be}_{\alpha\mu}(r)$, while the ellipsis contains the terms that will cancel. 
%
%\begin{align}
%    \frac{ig^4}{2}   h_1^{bedx}  \gt_{\alpha\beta\sigma\rho}^{(0)adxe} = \frac{3}{4} i\lambda^2 \delta^{ab} (g_{\alpha\beta}g_{\sigma\rho}\! - \! g_{\alpha\sigma}g_{\beta\rho}) \,.
%\end{align}

Finally, the contraction of 
$(d_{\s{2,5}})$ by $p^\nu$ activates  
the STI of \eq{BBQQw}, giving  
\begin{align}   \label{eq_d25}
    p^\nu (d_{\s{2,5}})^{abc}_{\alpha\mu\nu}  = & -\frac{ig^4}{2} f^{bcx}h_1^{xedm} \!\!\! \int_k \int_l \gt_{\alpha\beta\sigma\rho}^{(0)adme}\yt_\mu^{\rho\lambda}(l,q-l)  \yy_\lambda^{\sigma\beta}(s,k) + \cdots \,,
\end{align}
which is exactly $- f^{bce}(a_{\s{6}})^{ae}_{\alpha\mu}(q)$.

Once again, all the terms inside the ellipses in \2eqs{eq_d22_d24}{eq_d25}, 
cancel exactly against 
the terms coming from \mbox{$p^\nu\left[ (d_\s{2,1})+(d_\s{2,3})\right]_{\alpha\mu\nu}$}, 
leading to the second line in \eq{eq_2gl2}.

Thus,  the above considerations demonstrate the 
validity of \1eq{stiblocks} for $i=3$.

\subsection{Two-loop ghost sector ({\it fourth block})} 
\label{sec_2gh}

\begin{figure}[t]
    \centering
    \includegraphics[scale=0.52]{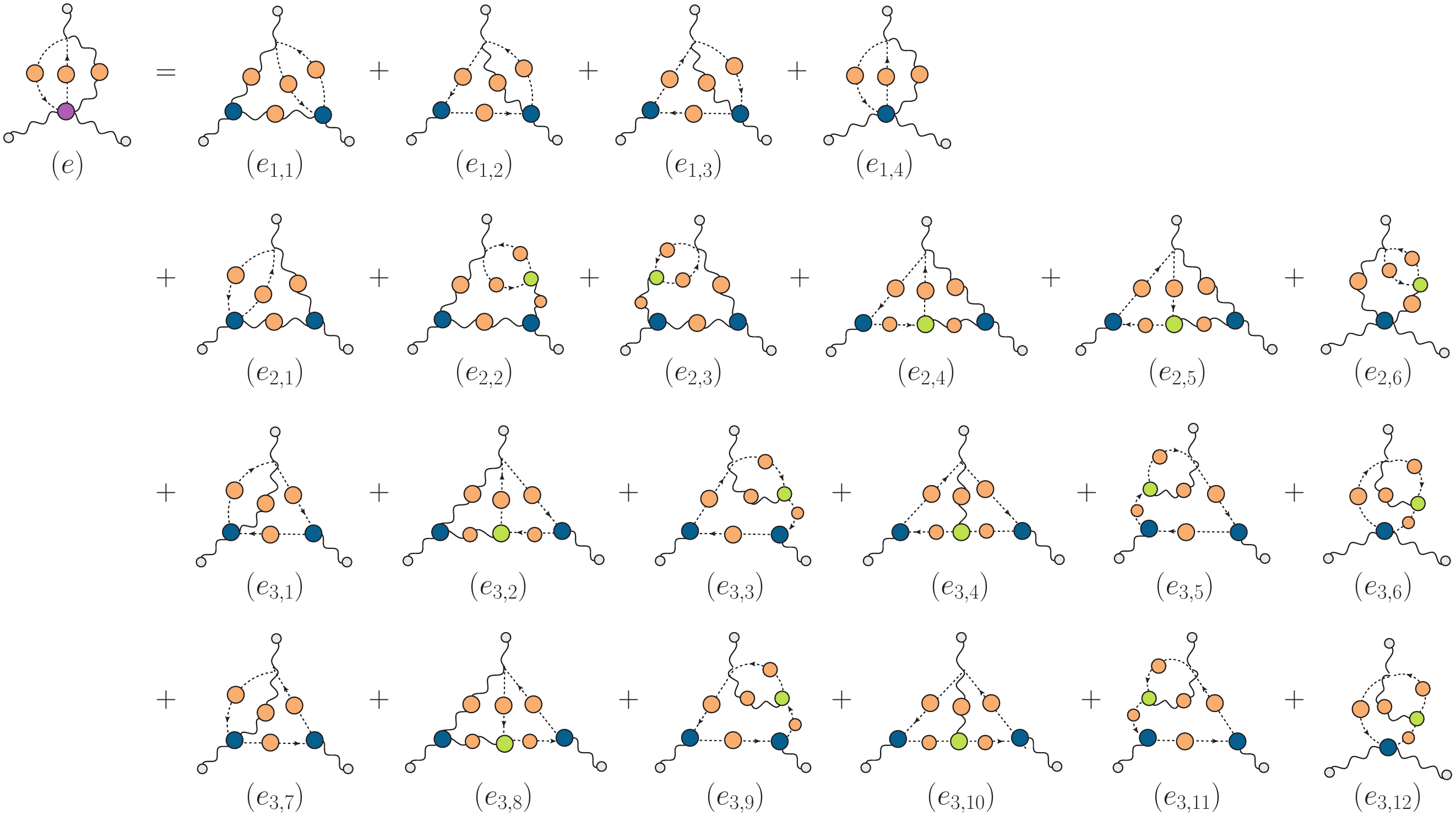}
    \caption{Contributions from the diagram $(e)$ in \fig{fig_bbbsde} after expanding the five-point kernel represented by the purple blob. This group is organized in the  three subsets $(e_{\s 1})$, $(e_{\s2})$, and $(e_{\s 3})$.}
    \label{fig_2gh}
\end{figure}

Finally, the two-loop dressed ghost graphs, given by the diagram $(e)$, enclosed by the green box in \fig{fig_bbbsde}, have twenty two contributions, depicted in \fig{fig_2gh}, which  
have been further separated into three subgroups as
\begin{align}   \label{eq_e}
    (e)^{abc}_{\alpha\mu\nu} 
    = \sum_{i=1}^{3} (e_{\s{i}})^{abc}_{\alpha\mu\nu} \,,
    %    + \sum_{i=2}^4\sum_{j=1}^{6} (e_{ij})^{abc}_{\alpha\mu\nu}(q,r,p) \,,
    %+ \sum_{j=1}^{6} (e_{3j})^{abc}_{\alpha\mu\nu}(q,r,p) 
    %+ \sum_{j=1}^{6} (e_{4j})^{abc}_{\alpha\mu\nu}(q,r,p)\,, 
\end{align}
with
\begin{align}   
\label{edecomp}
    (e_{\s{1}})^{abc}_{\alpha\mu\nu} = \sum_{j=1}^{4} (e_{\s{1,j}})^{abc}_{\alpha\mu\nu} \,,\qquad
    (e_{\s{2}})^{abc}_{\alpha\mu\nu} = \sum_{j=1}^{6} (e_{\s{2,j}})^{abc}_{\alpha\mu\nu}\,, \qquad 
    (e_{\s{3}})^{abc}_{\alpha\mu\nu} = \sum_{j=1}^{12} (e_{\s{3,j}})^{abc}_{\alpha\mu\nu} \,,
\end{align}
such that 
\begin{align}
    p^\nu(e_{\s{1}})^{abc}_{\alpha\mu\nu}\!&=\!f^{cae}(a_{\s 7})^{be}_{\alpha\mu}(r) \!-\!f^{bce}(a_{\s 7})^{ae}_{\alpha\mu}(q)\,, \nonumber \\
    p^\nu(e_{\s{2}})^{abc}_{\alpha\mu\nu}\!&=\!f^{cae}(a_{\s 8})^{be}_{\alpha\mu}(r)\!-\!f^{bce}(a_{\s 8})^{ae}_{\alpha\mu}(q) \,, \nonumber \\
    p^\nu (e_{\s{3}})^{abc}_{\alpha\mu\nu}\!&=\!f^{cae}\left[(a_{\s 9})\!+\!(a_{\s{10}})\right]^{be}_{\alpha\mu}(r)\!-\!f^{bce}\left[(a_{\s 9})\!+\!(a_{\s{10}})\right]^{ae}_{\alpha\mu}(q) \,. \label{eq_2gh}
\end{align}
The expressions for all the diagrams in \fig{fig_2gh}, 
together with the associated self-energy graphs, are given in the Appendix~\ref{sec_expressions}.

A close inspection of these expressions reveals that  
\begin{align}    
\label{eq_pe_1}
    &p^\nu (e_{\s{1,4}})^{abc}_{\alpha\mu\nu}=-f^{bce}(a_{\s 7})^{ae}_{\alpha\mu}(q) + \cdots  \,,   \qquad\quad   p^\nu (e_{\s{3,6}})^{abc}_{\alpha\mu\nu}= -f^{bce}(a_{\s 9})^{ae}_{\alpha\mu}(q) + \cdots  \,,\nonumber \\
    &p^\nu (e_{\s{2,6}})^{abc}_{\alpha\mu\nu}= -f^{bce}(a_{\s 8})^{ae}_{\alpha\mu}(q)  + \cdots  \,,    \qquad \quad p^\nu (e_{\s{3,12}})^{abc}_{\alpha\mu\nu}=
    -  f^{bce}(a_{\s{10}})^{ae}_{\alpha\mu}(q) + \cdots  \,, 
\end{align}
and  
\begin{align}       
\label{eq_pe_2} 
    p^\nu \left[(e_{\s{1,1}})\!+\!(e_{\s{1,2}})\!+\!(e_{\s{1,3}})\right]^{abc}_{\alpha\mu\nu}=&f^{cae}(a_{\s 7})^{be}_{\alpha\mu}(r)+ \cdots  \,, \nonumber \\
    p^\nu \left[(e_{\s{2,2}})\!+\!(e_{\s{2,4}})\right]^{abc}_{\alpha\mu\nu}=&f^{cae}(a_{\s 8})^{be}_{\alpha\mu}(r)+\cdots \,, \nonumber\\
    p^\nu \left[(e_{\s{3,3}})\!+\!(e_{\s{3,4}})\!+\!(e_{\s{3,9}})\!+\!(e_{\s{3,10}})\right]^{abc}_{\alpha\mu\nu}=& f^{cae}\left[(a_{\s 9})\!+\!(a_{\s{10}})\right]^{be}_{\alpha\mu}(r)+\cdots \,. 
\end{align}

%

%%%%%%%%%%%%%%%%%%%%%%%%%%%%%%%%%%%%%%%%%%%%%%%%%%%%%%%%%%%%%%%%%%%%%%
% Fig. 6 -  Skeleton expansion of diagram (e) -  five point kernel 3gluons-2 ghosts
%%%%%%%%%%%%%%%%%%%%%%%%%%%%%%%%%%%%%%%%%%%%%%%%%%%%%%%%%%%%%%%%%%%%%%%
%%%%%%%%%%%%%%%%%%%%%%%%%%%%%%%%%%%%%%%%%%%

It is then a matter of straightforward algebra 
to demonstrate that all terms contained in 
the ellipses of \2eqs{eq_pe_1}{eq_pe_2} cancel
against each other and with the other diagrams, as
\begin{align}
     p^\nu [(e_{\s{2,1}})\!+\!(e_{\s{2,3}})\!+\!(e_{\s{2,5}})]^{abc}_{\alpha\mu\nu}+ \cdots = 0 \,, \nonumber \\
     p^\nu [(e_{\s{3,1}})\!+\!(e_{\s{3,2}})\!+\!(e_{\s{3,3}})\!+\!(e_{\s{3,5}})\!+\!(e_{\s{3,7}})\!+\!(e_{\s{3,8}})\!+\!(e_{\s{3,11}})]^{abc}_{\alpha\mu\nu} + \cdots = 0\,,
\end{align}
leaving \eq{eq_2gh} 
as the final result. Thus, the 
validity of \1eq{stiblocks} for $i=4$ is confirmed. 

The final conclusion drawn 
from the analysis presented in  
subsections \ref{sec_1gl}, \ref{sec_1gh},   \ref{sec_2gl},  and \ref{sec_2gh} 
is that the block-wise
realization of the STI announced in subsection  
\ref{sec_gencon}, holds.  
Notice, in fact, that 
the validity of 
\eq{stiblocks} has been demonstrated for 
an {\it arbitrary} value of the gauge-fixing parameter $\xi_{\s Q}$.

\section{Abelian Ward identities with Background gluons}  
\label{sec_soft}

In this section we derive Abelian WIs 
from the STIs satisfied by the BFM vertices, 
and apply to them  
the text-book diagrammatic representation for the WIs 
known from QED~\cite{Itzykson:1980rh}. 
In addition, we demonstrate the 
block-wise realization of the WI that connects
the vertex \mbox{$\gh_{\alpha\mu\nu}(0,r,-r)$} 
with the derivative of ${\widehat\Delta}(r)$.  

%%%%%%%%%%%%%%%%%%%%%%%%%%%%%%%%%%%%%%%%%%%%%%%%%%%%%%%
% Fig. 7  Diagrammatic representation of WI for the Bcc vertex 
%%%%%%%%%%%%%%%%%%%%%%%%%%%%%%%%%%%%%%%%%%%%%%%%%%%%%%%%
\begin{figure}[t]
    \centering
    \includegraphics[scale=0.5]{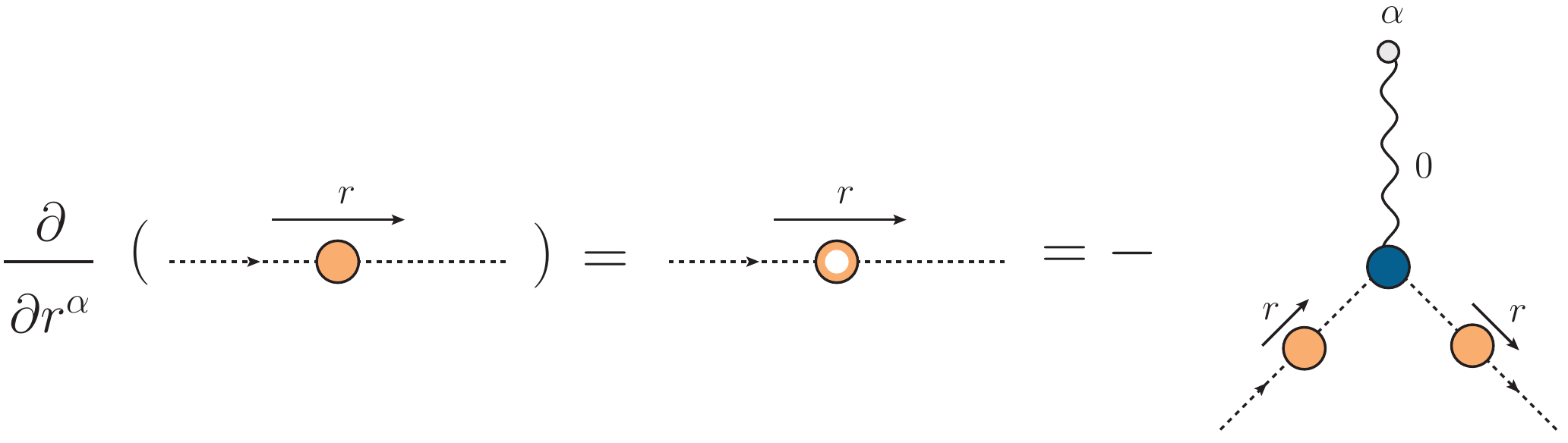}
    \caption{Diagrammatic representation of the WI for the \bcc\ vertex, where the derivative of the ghost propagator can be identified as the insertion of a zero-momentum background gluon leg. Here we define the notation of a perforated circle as being the derivative acting on the Green's function.}
    \label{fig_bccwi}
\end{figure}
%%%%%%%%%%%%%%%%%%%%%%%%%%%%%%%%%%%%%%%%%%

As is well-known in the context of Abelian gauge theories, 
such as spinor or scalar QED, the 
implementation of the 
limit $q\to 0$ of the Takahashi identity gives rise to the 
corresponding WI. 
In order to fix the ideas consider the latter theory, describing the interaction of a photon with a complex scalar, where the full photon-scalar vertex $\Gamma_\mu(r,p,q)$ 
satisfies the Abelian STI (Takahashi identity) 
\be 
q^\mu \Gamma_\mu(r,p,q) = {\cal D}^{-1}(p) - {\cal D}^{-1}(r),
\label{WIAphi2}
\ee
with ${\cal D}(p)$ denoting the fully dressed propagator of the scalar field. Then, the  
standard WI is determined by expanding both sides of \1eq{WIAphi2} around $q=0$, and equating the linear terms. Specifically, this procedure 
yields 
\be
\Gamma_\mu(r,-r,0) =  \frac{\partial \,{\cal D}^{-1}(r)}{\partial r^\mu}\,,
\label{wisqed}
\ee
or, equivalently, 
\be
\frac{\partial \,{\cal D}(r)}{\partial r^\mu} = -
{\cal D}(r) \Gamma_\mu(r,-r,0) {\cal D}(r)\,.
\label{wisqed2}
\ee
The version of the WI given in \1eq{wisqed2}
admits the text-book diagrammatic interpretation: the derivative of the propagator  ${\cal D}(r)$
is equivalent to the insertion of a 
zero-momentum photon in it~\cite{Itzykson:1980rh}. 

It turns out  
that the Abelian STIs satisfied by the BFM 
three-point functions give rise to 
WIs completely analogous to that of \1eq{wisqed2}, which 
admit the same diagrammatic 
interpretation given above, but now in terms of  
zero-momentum insertions of a background gluon.

The simplest case is that of the ghost-gluon vertex 
$\gt_\mu(r,p,q)$, whose WI is identical to 
that of \2eqs{wisqed}{wisqed2}, after the replacement  
$\Gamma_\mu \to \gt_\mu$ and ${\cal D}\to D$, \ie 
\be
\gt_\mu(r,-r,0) =  \frac{\partial \,{D}^{-1}(r)}{\partial r^\mu} \,
\Longrightarrow \,
\frac{\partial \,{D}(r)}{\partial r^\mu} = -
{ D}(r) \gt_\mu(r,-r,0) {D}(r)\,;
\label{wisghgl}
\ee
the corresponding diagrammatic representation is shown in 
\fig{fig_bccwi}.

\begin{figure}[t]
    \centering
    \includegraphics[scale=0.5]{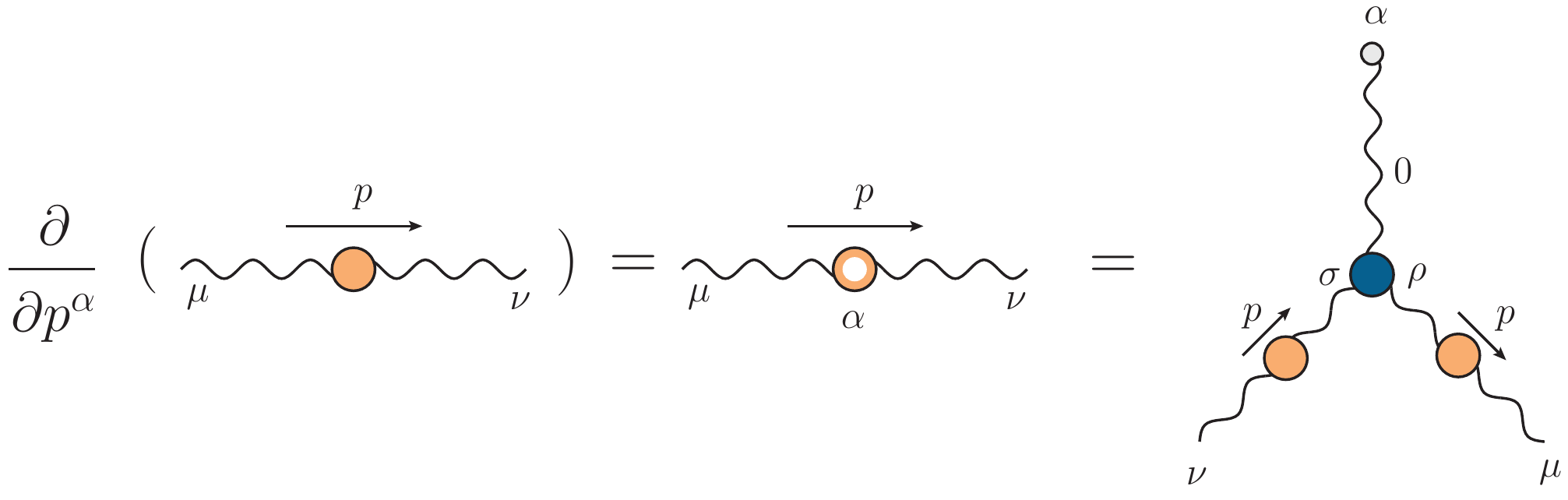}
    \caption{Diagrammatic representation of the WI for the \bqq\ vertex,  where the derivative of the gluon propagator can be identified as the insertion of a zero-momentum background gluon leg.}
    \label{fig_bqqwi}
\end{figure}

Turning to the case of the BQQ vertex $\widetilde{\Gamma}_{\alpha\mu\nu}(q,r,p)$, 
it is rather straightforward to deduce from the  
STI of \1eq{BQQw} the corresponding WI, namely 
\begin{equation}
 \widetilde{\Gamma}_{\alpha\mu\nu}(0,-p,p) =- \frac{\partial \Delta^{-1}_{\mu\nu}(p)}{\partial p^\alpha}\,
\Longrightarrow \,
\frac{\partial \Delta^{\mu\nu}(p)}{\partial p^\alpha} = 
\Delta^{\mu\rho}(p)\,\widetilde{\Gamma}_{\alpha\rho\sigma}(0,-p,p) \,\Delta^{\nu\sigma}(p) \,;
\label{wi3g}
\end{equation}
the last relation is diagrammatically depicted in 
\fig{fig_bqqwi}.
Note that the above WI, when applied at tree level,  
reproduces from \1eq{invBQprop}
the expression for $\gt_{\alpha\mu\nu}^{(0)}(q,r,p)$ 
given in \1eq{BQQ0}, capturing correctly its
dependence on the 
gauge-fixing parameter $\xi_{\s Q}$. 

%%%%%%%%%%%%%%%%%%%%%%%%%%%%%%%%%%%%%%%%%%
%Fig. 8  Diagrammatic representation of WI for the BQQ vertex 
%%%%%%%%%%%%%%%%%%%%%%%%%%%%%%%%%%%%%%%%%%
%%%%%%%%%%%%%%%%%%%%%%%%%%%%%%%%%%%%%%%%%%

\begin{comment}
    Here there was the WI for the BQQQ, I changed for the BBcc, which is going to be useful at some point.

    We next focus our attention on the WIs satisfied by BFM vertices with 
    more than three incoming fields.
    As a concrete example, consider the vertex 
    $BQQQ$, whose STI when contracted with respect to the 
    momentum carried by the background gluon is given 
    by \1eq{BQQQw}.  
    Expanding both sides of \1eq{BQQQw} around $q=0$, and using the Jacobi identity to eliminate the zeroth order term on the r.h.s., we obtain the WI 
    %
    \begin{equation}
    \widetilde{\Gamma}^{abmn}_{\alpha\beta\mu\nu}(0,r,p,-r-p) =  \bigg(f^{abx}f^{xnm}\frac{\partial}{\partial r^\alpha} + f^{amx}f^{xbn}\frac{\partial}{\partial p^\alpha}\bigg)
    \Gamma_{\beta\mu\nu}(r,p,-r-p).
    \label{wibqqq}
    \end{equation}
    %
    Exactly analogous expressions may be deduced for higher point Green's
    functions; for a formal derivation of the STI satisfied by a general vertex of the form $BQ^n$, see Appendix \ref{sec_derivation}.
\end{comment}

We next focus our attention on the WIs satisfied by BFM vertices with 
more than three incoming fields. As a concrete example, consider the vertex \bbcc;  when contracted with respect to the momentum carried by one of the background gluons, it satisfies the STI   given by \1eq{BBccw}. Expanding both sides of \1eq{BBccw} around $q=0$, and using the Jacobi identity to eliminate the zeroth order term on the r.h.s., we obtain the WI 
\begin{equation}
    \gt^{abmn}_{\mu\nu}(0,-p-t,p,t) = \left(f^{amx}f^{nbx} \frac{\partial}{\partial p^\mu} + f^{anx}f^{bmx} \frac{\partial}{\partial t^\mu} \right) \gt_\nu(p,t,-p-t) \,.
    \label{wibbcc}
\end{equation}
Exactly analogous expressions may be deduced for higher point Green's 
functions; for a formal derivation of the STI satisfied by a general vertex of the form ${\rm BQ}^n$, see Appendix \ref{sec_derivation}.

Now we want to explore the block-wise realization of the WI of the \bbb\ vertex for the case of the one-loop ghost group, which satisfies the 
STI of \1eq{stiblocks} for $i=2$, 
or, equivalently, \eq{eq_1gh0}. 
In the soft-gluon limit, we obtain simply
\begin{align} 
    \gh^{(2)}_{\alpha\mu\nu}(q,-q,0)=\frac{\partial \ph^{(2)}_{\alpha\mu}(q)}{\partial q^\nu} \,, 
\end{align}
or, in terms of diagrams
\begin{align}   \label{eq_WIc3}
    (c_{\s{3}})_{\alpha\mu\nu}(q,-q,0)=\frac{\partial(a_{\s{3}})_{\alpha\mu}(q)}{\partial q^\nu} \,.
\end{align}
In arriving at \1eq{eq_WIc3} 
we have used that 
$(a_{\s{4}})$ is $q$-independent, 
and that, 
in the soft-gluon limit, 
$(c_{\s{1}}) = (c_{\s{2}}) =0$ 
(see Sec.~\ref{sec_1gh}).

\begin{figure}
    \centering
    \includegraphics[scale=0.5]{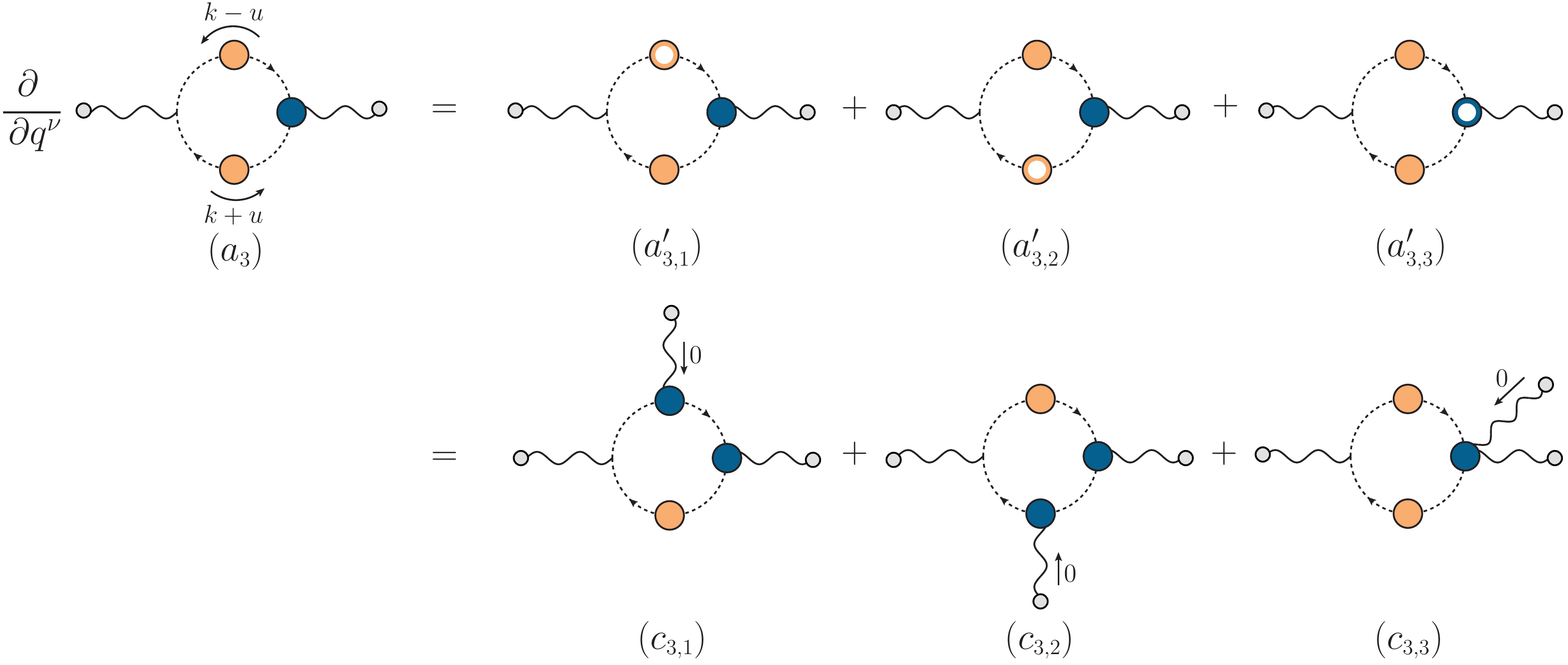}
    \caption{The diagrammatic 
    representation of the differentiation of  the 
    graph $(a_{\s{3}})$ with respect to $q^{\nu}$.  The effect of differentiating the ghost propagator (dressed background ghost-gluon vertex) is the insertion of a zero-momentum background gluon leg in the propagator (vertex).}
    \label{der_a3}
\end{figure}

To prove \eq{eq_WIc3}, we first
symmetrize the process of 
differentiation of 
$(a_{\s{3}})_{\alpha\mu}$
by shifting the loop momentum  
($k \rightarrow \msym-k$, with $\msym = q/2$), to get 

\begin{align}   \label{eq_a3der}
\frac{\partial(a_{\s{3}})_{\alpha\mu}(q)}{\partial q^\nu} = 
(a^{\prime}_{\s{3,1}})_{\alpha\mu\nu}(q) 
+ (a^{\prime}_{\s{3,2}})_{\alpha\mu\nu}(q) 
+
(a^{\prime}_{\s{3,3}})_{\alpha\mu\nu}(q) 
    \,,
\end{align}
with 
\bea 
(a^{\prime}_{\s{3,1}})_{\alpha\mu\nu}(q) &=&-2\lambda  \int_k k_\alpha \left[ \frac{\partial}{\partial q^\nu}D(k-\msym)\right] D(k+\msym)  \gt_\mu(k+\msym,\msym-k,-q)\,, 
\nonumber\\
(a^{\prime}_{\s{3,2}})_{\alpha\mu\nu}(q) &=& 
-2\lambda  \int_k k_\alpha D(k-\msym) \left[ \frac{\partial}{\partial q^\nu}D(k+\msym)\right]  \gt_\mu(k+\msym,\msym-k,-q)\,,\nonumber\\
(a^{\prime}_{\s{3,3}})_{\alpha\mu\nu}(q) &=& -2\lambda  \int_k k_\alpha D(k-\msym) D(k+\msym) \left[ \frac{\partial}{\partial q^\nu} \gt_\mu(k+\msym,\msym-k,-q) \right]\,;   \label{eq_a3limit}
\eea 
the last three 
contributions are depicted graphically 
in the first line of \fig{der_a3}. 

Next, 
for the terms $(a^{\prime}_{\s{3,1}})_{\alpha\mu\nu}(q)$ 
and $(a^{\prime}_{\s{3,2}})_{\alpha\mu\nu}(q)$ we use 
\eq{wisghgl} to write 
\begin{align}  \label{eq_a3limit_fin}
    (a^{\prime}_{\s{3,1}})_{\alpha\mu\nu}(q)\!=\!-&\lambda \!\! \int_k k_\alpha  \left[ D(k\!-\!\msym)\gt_\nu(k\!-\!\msym,\msym\!-\!k,0) D(k\!-\!\msym)\right] \!D(k\!+\!\msym) \gt_\mu(k\!+\!\msym,\msym\!-\!k,-q) \,, \nonumber \\
    (a^{\prime}_{\s{3,2}})_{\alpha\mu\nu}(q)\!=\!-&\lambda \!\! \int_k k_\alpha  D(k\!-\!\msym) \left[ D(k\!+\!\msym) \gt_\nu(k\!+\!\msym,-\msym\!-\!k,0) D(k\!+\!\msym) \right]   \gt_\mu(k\!+\!\msym,\msym\!-\!k,-q) \,. 
\end{align}
A direct comparison of these last expressions  with the contributions to \mbox{$(c_{\s{3}})_{\alpha\mu\nu}(q,-q,0)$} in \eq{eq_c3} [for $(q,r,p) \to (q,-q,0)$] allows one to establish that  
\begin{align}
    (a^{\prime}_{\s{3,1}})_{\alpha\mu\nu}(q)= (c_{\s{3,1}})_{\alpha\mu\nu}(q,-q,0)\,, \qquad (a^{\prime}_{\s{3,2}})_{\alpha\mu\nu}(q)= (c_{\s{3,2}})_{\alpha\mu\nu}(q,-q,0)\,.
\end{align}
%
%%%%%%%%%%%%%%%%%%%%%%%%%%%%%%%%%%%%%%%%%%%%%%%%%%%%%%%%%%%%%%%%%
%  Fig. 9  - Effect of acting the derivative on the ghost loop
%%%%%%%%%%%%%%%%%%%%%%%%%%%%%%%%%%%%%%%%%%%%%%%%%%%%%%%%%%%%%%%%%
%%%%%%%%%%%%%%%%%%%%%%%%%%%%%%%%%%%%%%%%%%%%%%%%%%%%%

Consider finally the $(c_{\s{3,3}})^{abc}_{\alpha\mu\nu}$ in 
\eq{eq_cs}; setting $(q,r,p) \to (q,-q,0)$, shifting $k \rightarrow \msym-k$, 
and employing \eq{wibbcc}, we get 
\begin{align}
    (c_{\s{3,3}})^{abc}_{\alpha\mu\nu}(q,-q,0) &= 2g^2 f^{eda} \int_k k_\alpha D(k-\msym) D(k+\msym)  \gh_{\mu\nu}^{bcde}(-q,0,k+\msym,\msym-k) \,, \nonumber \\
    &= -\frac{\lambda}{2}\!\! \int_k  \!\! k_\alpha D(k\!-\!\msym) D(k\!+\!\msym) \!\! \left( \! \frac{\partial}{\partial (k+\msym)^\nu}\! +\! \frac{\partial}{\partial (\msym-k)^\nu} \!\!\right)\! \gt_\mu(k\!+\!\msym,\msym\!-\!k,-q)  \,, \nonumber \\
    &= -2\lambda \int_k k_\alpha D(k-\msym) D(k+\msym) \frac{\partial}{\partial q^\nu}  \gt_\mu(k+\msym,\msym-k,-q)\,,
\end{align}
and therefore
\begin{align}
    (a^{\prime}_{\s{3,3}})_{\alpha\mu\nu}(q)= (c_{\s{3,3}})_{\alpha\mu\nu}(q,-q,0)\,,  
\end{align}
which completes the proof of 
\eq{eq_WIc3}. The interpretation of the previous steps in terms of 
background-gluon 
insertions is given in the second line of \fig{der_a3}.

\section{Discussion and Conclusions}
\label{sec_disc}

It has been known for some time~\cite{Aguilar:2006gr} that the transversality 
of the background self-energy is enforced in a special 
way, namely independently for each one of the four subsets 
(blocks) of diagrams comprising the corresponding SDE. 
In the present work we have shown that 
the Abelian STI of the background three-gluon vertex  
is also realized according to the exact same pattern, 
at the level of the corresponding SDE: 
the momentum contraction of each subset of vertex  
diagrams generates the difference of the 
corresponding self-energy subsets.

The demonstration of this property has been carried out at the level of the 
fully dressed Feynman diagrams that comprise the 
relevant SDEs. In particular, the contraction 
of all three-gluon vertex diagrams by the 
appropriate momentum triggers STIs satisfied by the vertices and the kernels 
embedded in them, giving rise to crucial 
rearrangements and cancellations, which are 
implemented algebraically, with no need to resort  
to any integrations. 
Note that the extensive reorganization of diagrams observed here 
has been first identified in the context of the pinch technique~\cite{Cornwall:1981zr,Papavassiliou:1989zd,Pilaftsis:1996fh,Binosi:2002ft,Binosi:2009qm}, 
where the ``gauge-invariant'' three-gluon vertex was first studied 
at the one-loop level~\cite{Cornwall:1989gv,Hashimoto:1994ct,Binger:2006sj}. 
Evidently, it would be particularly interesting to 
explore the origin of the block-wise STIs at a 
formal level, and establish its validity by means of the  
Batalin-Vilkovisky functional machinery ~\cite{Batalin:1977pb,Batalin:1983ggl,Binosi:2002ez,Binosi:2009qm,Binosi:2012st,Binosi:2013cea} .   

It is natural to conjecture that the 
STI of the background four-gluon vertex, B${}^4$, given by~\cite{Papavassiliou:1992ia,Hashimoto:1994ct}
\bea 
q^\mu \gb^{mnrs}_{\mu\alpha\beta\gamma}(q,r,p,t) &=& f^{mse}f^{ern} \gb_{\alpha\beta\gamma}(r,p,q+t) + f^{mne}f^{esr}\gb_{\beta\gamma\alpha}(p,t,q+r)
\nonumber \\
&+& f^{mre}f^{ens} \gb_{\gamma\alpha\beta}(t,r,q+p)\,,
\label{4g_STI}
\eea 
is realized according to the same block-wise pattern described above. 
A diagrammatic demonstration along the lines presented in this work appears to be quite feasible, and would 
give further support to the notion that the STI of  
any B${}^n$-type of vertex is enforced in this  
characteristic manner.

Some of the results presented in Sec.~\ref{sec_soft}
may be used in order to explore the numerical 
impact of certain truncations or approximations, in the 
spirit of the recent study presented in~\cite{Aguilar:2022wsh}.
For example, the  
equality shown in \fig{der_a3} will be distorted 
if the vertex BB${\rm \bar c} \rm c$ were to be replaced by its tree level 
value, given by \1eq{BBcc0}. The amount of discrepancy induced 
between the two sides of this equation is a quantitative indicator of the 
veracity of such an approximation.

Throughout the present analysis we have assumed that the BBB vertex 
does not contain irregularities in the form of massless poles. However, 
as has been shown in detail in a series of studies, 
the emergence of a dynamical gluon mass~\cite{Cornwall:1981zr}
through the operation of the Schwinger mechanism~\cite{Schwinger:1962tn,Schwinger:1962tp}
hinges on the inclusion of 
longitudinally coupled massless poles in the 
fundamental vertices of the theory
~\cite{Jackiw:1973tr,Eichten:1974et,Aguilar:2008xm,Aguilar:2011xe,Ibanez:2012zk,Aguilar:2016vin,Eichmann:2021zuv}
Quite importantly: ({\it a}) the 
STIs satisfied by the vertices 
are resolved with the nontrivial participation of these poles, 
and ({\it b})
in the soft-gluon limit, the associated WIs are {\it displaced} 
by an amount controlled by the corresponding pole residues~\cite{Aguilar:2016vin,Aguilar:2021uwa,Aguilar:2022thg}. 
In particular, ongoing research reveals that the STIs impose 
stringent conditions on the pole content of 
the three-gluon vertex, which must, at the same time, 
be dynamically realized. 
The treatment of this problem within the BFM 
(\ie at the level of the BBB rather than the QQQ vertex)  
eliminates structures originating from the ghost-sector of the theory, 
which tend to complicate and obscure the underlying physical  
picture. We expect that the completion of this study will shed light on  
the question of how symmetry-induced constraints      
are dynamically enforced at the level the corresponding SDEs. 

%%%%%%%%%%%%%%%%%%%%%%%%%%%%%%%%%%%%%%%%%%%%%%%%

\section{Acknowledgments}
\label{sec:acknowledgments}
The work of  A.~C.~A. and B.~M.~O. are supported by the CNPq grants \mbox{307854/2019-1} and 141409/2021-5, respectively.
A.~C.~A also acknowledges financial support from  project 464898/2014-5 (INCT-FNA).
M.~N.~F. and J.~P. are supported by the Spanish MICINN grant PID2020-113334GB-I00. M.~N.~F. acknowledges financial support from Generalitat Valenciana through contract CIAPOS/2021/74. J.~P. also acknowledges  
funding from the regional Prometeo/2019/087 from the Generalitat Valenciana.

\appendix

\section{Derivation of Abelian STIs} 
\label{sec_derivation} 

In this Appendix we 
employ the Batalin-Vilkovisky formalism~\cite{Batalin:1977pb,Batalin:1983ggl,Binosi:2002ez,Binosi:2009qm,Binosi:2012st,Binosi:2013cea} to derive the Abelian 
STI satisfied by the generic vertex ${\rm BQ}^n$
when contracted by the momentum carried by the gluon $\rm B$.

We start with the WI functional, given by~\cite{Binosi:2009qm}
\begin{align}
\hspace{-0.3cm}
    W \!= \!\!\int \!\! d^4x \!\left[ \delta_\vartheta Q^{x,\mu}(x) \frac{\delta \Gamma}{\delta Q_\mu^x(x)} \!+\!
    \delta_\vartheta B^{x,\mu}(x) \frac{\delta \Gamma}{\delta B_\mu^x(x)} \!+\! 
    \delta_\vartheta c^x(x) \frac{\delta \Gamma}{\delta c^x(x)} \!+\!
    \delta_\vartheta \bar{c}^x(x) \frac{\delta \Gamma}{\delta \bar{c}^x(x)} \right]\!\!=0\,,  
    \label{WI_func}
\end{align}
where $\vartheta^a$ are the local infinitesimal parameters which correspond to the SU(3) generators $t^a$, and play the role of the ghost field. $\Gamma$ in \1eq{WI_func} is the ``reduced'' effective action, defined as the full effective action without the gauge-fixing term~\cite{Binosi:2002ez,Binosi:2009qm}.
Consequently, the Green's functions obtained from $\Gamma$ will be missing the corresponding gauge-dependent contribution at tree level.
Finally, the gauge transformations of the fields are given by
\begin{align}
    \delta_\vartheta Q_\mu^x=gf^{xdc}Q_\mu^d \vartheta^c\,, \qquad &\delta_\vartheta B_\mu^x=\partial_\mu \vartheta^x + gf^{xdc}B_\mu^d \vartheta^c\,,  \\
    \delta_\vartheta c^x=-gf^{xdc}c^d \vartheta^c\,, \qquad &\delta_\vartheta \bar{c}^x=-gf^{xdc}\bar{c}^d \vartheta^c\,. \nonumber 
\end{align}

To obtain the background Abelian STIs the first step is to differentiate the functional $W$ with respect to the parameter $\vartheta^a(x)$, furnishing
\begin{align} \label{eq_start}
    \frac{\delta W}{\delta \vartheta^a(x)}= gf^{eda}Q_\mu^d(x) \Gamma_{Q_\mu^e}(x) + \partial_\mu \Gamma_{B_\mu^a}(x) =0 \,,
\end{align}
where we have already set to zero the vacuum expectation values (VEVs) of the ghost, antighost, and background gluon fields\footnote{In the end of the procedure all of the VEVs are set to zero. Since the ${\rm BQ}^n$ vertex has only one external $\rm B$ and no external ghost and antighost fields these VEVs can be set to zero from the outset.}. Moreover, we introduce the shorthand notation for functional derivatives
\be 
\Gamma_{\phi_1 \phi_2 \ldots \phi_n}(x_1, x_2, \ldots, x_n) := \frac{\delta^{n} \Gamma}{\delta \phi_1(x_1) \delta \phi_2(x_2) \ldots \delta \phi_n(x_n) } \,, \label{der_def}
\ee 
where $\phi_i(x_i)$ denotes a generic field.

The STIs of interest are then obtained by differentiating \1eq{eq_start} $n$ times with respect to the quantum gluon. Note, in particular, that the functional derivatives of the term $\partial_\mu \Gamma_{B_\mu^a}(x)$ in \1eq{eq_start} generate divergences such as $\partial_\mu \Gamma_{B_\mu^a Q_{\nu_{\s{1}}}^{b_{\s{1}}}\ldots Q_{\nu_{\s{n}}}^{b_{\s{n}}}}(x,y_1, \ldots, y_n)$ which, after Fourier transformation, result in the typical l.h.s. of the Abelian STIs, \ie a Green's function contracted with a background gluon momentum.

To fix the ideas, let us consider as two special cases the STIs for the BQ and BQQ functions. 

Differentiating \1eq{eq_start} with respect to $Q_{\nu_1}^{b_{\s{1}}}(y_1)$ we obtain
\begin{align}   \label{eq_start1}
    \frac{\delta^2 W}{\delta Q_{\nu_1}^{b_{\s{1}}}(y_1)\delta \vartheta^a(x)}\!\!=& gf^{eb_{\s{1}}a} \delta(x\!-\!y_1) \Gamma_{\!Q_{\nu_1}^e}\!\!(x) \!+\! gf^{eda}Q_\mu^d(x) \Gamma_{\!Q_{\nu_1}^{b_{\s{1}}}Q_\mu^e}\!(y_1,x)\!+
    \!\partial_\mu  \Gamma_{\!B_\mu^aQ_{\nu_1}^{b_{\s{1}}}}\!(x,y_1) \! =\!0 \,.
\end{align}
Setting the gluon field $Q=0$, and the one-point function $\Gamma_{\!Q_{\nu_1}^e}\!\!(x)=0$, we obtain 
\begin{align}   
     \partial_\mu \Gamma_{B_\mu^aQ_{\nu_1}^{b_{\s{1}}}}(x,y_1)  =0  \,, \label{BQ_STI}
\end{align}
where $\Gamma_{B_\mu^aQ_{\nu_1}^{b_{\s{1}}}}(x,y_1)$ is the inverse BQ propagator, with its $1/\xi_Q$ term removed. In momentum space notation, \1eq{BQ_STI} becomes
\be 
q^\mu\left[ q^2P_{\mu\nu}(q) + i{{\widetilde\Pi}_{\mu\nu}(q)}\right] = 0 \quad \Longrightarrow \quad q^\mu{{\widetilde\Pi}_{\mu\nu}(q)} = 0 \,,
\ee 
expressing the  exact transversality of the BQ self-energy.

Then, an additional  differentiation of  \1eq{eq_start1} with respect to $Q_{\nu_2}^{b_2}(y_2)$ yields
\begin{align}   \label{eq_qqv}
    \frac{\delta^3 W}{\delta Q_{\nu_2}^{b_{\s{2}}}(y_1)\delta Q_{\nu_1}^{b_{\s{1}}}(y_1)\delta \vartheta^a(x)}= &gf^{eb_{\s{1}}a} \delta(x-y_1) \Gamma_{Q_{\nu_1}^eQ_{\nu_2}^{b_{\s{2}}}}(x,y_2) + gf^{eb_{\s{2}}a}\delta(x-y_2)\Gamma_{Q_{\nu_1}^{b_{\s{1}}}Q_{\nu_2}^e}(y_1,x) \nonumber \\
    + gf^{eda}Q_\mu^d(x)& \Gamma_{ Q_{\nu_1}^{b_{\s{1}}}Q_{\nu_2}^{b_{\s{2}}}Q_\mu^e}(y_1,y_2,x)  + \partial_\mu  \Gamma_{B_\mu^aQ_{\nu_1}^{b_{\s{1}}}Q_{\nu_2}^{b_{\s{2}}}}(x,y_1,y_2)  =0   \,.
\end{align}
At this point, by setting all the fields to zero we obtain the Abelian STI for the BQQ vertex in configuration space, namely
\begin{align}
    \partial_\mu \Gamma_{B_\mu^aQ_{\nu_1}^{b_{\s{1}}}Q_{\nu_2}^{b_{\s{2}}}}(x,y_1,y_2) \!=\! -gf^{eab_{\s{1}}} \delta(x\!-\!y_1) \Gamma_{Q_{\nu_1}^eQ_{\nu_2}^{b_{\s{2}}}}(x,y_2)\!-\! gf^{eab_{\s{2}}}\delta(x\!-\!y_2)\Gamma_{Q_{\nu_1}^{b_{\s{1}}}Q_{\nu_2}^e}(y_1,x) \,. \label{BQQ_STI_conf}
\end{align}
Now, Fourier transforming the above equation to momentum space leads to
\begin{align}
    q^\mu \Gamma_{B_\mu^aQ_{\nu_1}^{b_{\s{1}}}Q_{\nu_2}^{b_{\s{2}}}}(q,r,p) = gf^{a b_{\s{1}}b_{\s{2}}} \left[ p^2P_{\mu\nu}(p) + {\Pi_{\mu\nu}(p)}\right]- gf^{ab_{\s{1}}b_{\s{2}}}\left[ r^2P_{\mu\nu}(r) + {\Pi_{\mu\nu}(r)}\right] \,, \label{BQQ_STI_mom}
\end{align}
which, with the definition of \1eq{3vert}, becomes \eq{BQQw}. The derivation of the STI for the BBB vertex, given in \eq{eq_WTIBBB}, is completely analogous.

Next, we prove that the STI of the ${\rm BQ}^n$ vertex is given by
\begin{align} \label{eq_bqn}
    &-\partial_\mu \Gamma_{B_\mu^aQ_{\nu_1}^{b_{\s{1}}}Q_{\nu_2}^{b_{\s{2}}}\cdots Q_{\nu_n}^{b_n}}(x,y_1,y_2, \cdots, y_n) = gf^{eab_{\s{1}}} \delta(x-y_1) \Gamma_{Q_{\nu_1}^eQ_{\nu_2}^{b_{\s{2}}}Q_{\nu_3}^{b_{\s{3}}}\cdots Q_{\nu_n}^{b_n}}(x,y_2,y_3, \cdots, y_n) \nonumber \\
    &\hspace{4cm}+ gf^{eab_{\s{2}}}\delta(x-y_2)\Gamma_{Q_{\nu_1}^{b_{\s{1}}}Q_{\nu_2}^eQ_{\nu_3}^{b_{\s{3}}}\cdots Q_{\nu_n}^{b_n}}(y_1,x,y_3,\cdots,y_n) \nonumber  +  \cdots \\
    &\hspace{4cm}+ gf^{eab_n}\delta(x-y_n)\Gamma_{Q_{\nu_1}^{b_{\s{1}}}Q_{\nu_2}^{b_{\s{2}}}\cdots Q_{\nu_{n-1}}^{b_{n-1}}Q_{\nu_n}^{e}}(y_1,y_2,\cdots,y_{n-1},x) \,. 
\end{align}
To that end, we first differentiate \eq{eq_qqv} $n-2$ times. This procedure yields
\begin{align} 
\label{eq_qnv}
   &-\partial_\mu \Gamma_{B_\mu^aQ_{\nu_1}^{b_{\s{1}}}Q_{\nu_2}^{b_{\s{2}}}\cdots Q_{\nu_n}^{b_n}}(x,y_1,y_2, \cdots, y_n) 
   = gf^{eab_{\s{1}}} \delta(x-y_1) \Gamma_{Q_{\nu_1}^eQ_{\nu_2}^{b_{\s{2}}}Q_{\nu_3}^{b_{\s{3}}}\cdots Q_{\nu_n}^{b_n}}(x,y_2,y_3, \cdots, y_n) \nonumber \\
    &\hspace{3cm}+ gf^{eab_{\s{2}}}\delta(x-y_2)\Gamma_{Q_{\nu_1}^{b_{\s{1}}}Q_{\nu_2}^eQ_{\nu_3}^{b_{\s{3}}}\cdots Q_{\nu_n}^{b_n}}(y_1,x,y_3,\cdots,y_n) \nonumber \\
    &\hspace{3cm}+ gf^{eax}\left\{ \frac{\delta^{n-2}}{\delta Q_{\nu_n}^{b_n}(y_n) \cdots \delta Q_{\nu_3}^{b_{\s{3}}}(y_3)} \left[Q_\mu^x(x)\Gamma_{Q_{\nu_1}^{b_{\s{1}}}Q_{\nu_2}^{b_{\s{2}}} Q_{\mu}^{e}}(y_1,y_2,x) \right] \right\}_{Q\rightarrow 0}\!\!\!\,. 
\end{align}

Clearly, to demonstrate \1eq{eq_bqn} we need to prove that
\begin{align}   
\label{eq_curly}
    gf^{eax}&\left\{ \frac{\delta^{n-2}}{\delta Q_{\nu_n}^{b_n}(y_n) \cdots \delta Q_{\nu_3}^{b_{\s{3}}}(y_3)}  \left[Q_\mu^x(x)\Gamma_{Q_{\nu_1}^{b_{\s{1}}}Q_{\nu_2}^{b_{\s{2}}} Q_{\mu}^{e}}(y_1,y_2,x) \right]\right\}_{Q\rightarrow 0}= \nonumber\\
    &\qquad \qquad \qquad \qquad \qquad 
    gf^{eab_{\s{3}}}\delta(x-y_3)\Gamma_{Q_{\nu_1}^{b_{\s{1}}}Q_{\nu_2}^{b_{\s{2}}}Q_{\nu_3}^{e}Q_{\nu_4}^{b_4}\cdots Q_{\nu_n}^{b_n}}(y_1,y_2,x,y_4,\cdots,y_n) +  \cdots \nonumber \\
    &\qquad \qquad \qquad \qquad \qquad +gf^{eab_n}\delta(x-y_n)\Gamma_{Q_{\nu_1}^{b_{\s{1}}}Q_{\nu_2}^{b_{\s{2}}}\cdots Q_{\nu_{n-1}}^{b_{n-1}}Q_{\nu_n}^{e}}(y_1,y_2,\cdots,y_{n-1},x) \nonumber \\
    & \qquad \qquad \qquad \qquad \qquad  + gf^{eax}\left[ Q_\mu^x(x)\Gamma_{Q_{\nu_1}^{b_{\s{1}}}Q_{\nu_2}^{b_{\s{2}}}\cdots Q_{\nu_{n}}^{b_{n}}Q_{\mu}^{e}}(y_1,y_2,\cdots,y_{n},x) \right]_{Q\rightarrow0}   \,. 
\end{align}
The proof proceeds by induction. First, it is clear that \1eq{eq_curly} holds for $n = 3$. Indeed, in this case one has to take a single derivative
\begin{align} 
&gf^{eax}\!\left\{\frac{\delta}{\delta Q_{\nu_{\s{3}}}^{b_{\s{3}}}(y_3)}\! \left[Q_\mu^x(x)\Gamma_{Q_{\nu_1}^{b_{\s{1}}}Q_{\nu_2}^{b_{\s{2}}} Q_{\mu}^{e}}(y_1,y_2,x) \right]\right\}_{Q\rightarrow 0} \!\!\!= gf^{eab_{\s{3}}}\delta(x-y_3)\Gamma_{Q_{\nu_1}^{b_{\s{1}}}Q_{\nu_2}^{b_{\s{2}}}Q_{\nu_3}^{e}}(y_1,y_2,x)\nonumber \\
 &\hspace{6cm}+ gf^{eax}\left[ Q_\mu^x(x)\Gamma_{Q_{\nu_1}^{b_{\s{1}}}Q_{\nu_2}^{b_{\s{2}}} Q_{\nu_{3}}^{b_{3}}Q_{\mu}^{e}}(y_1,y_2,y_3,x) \right]_{Q\rightarrow0} \,,
\end{align}
which is \1eq{eq_curly} for $n = 3$.

Then, assume that \1eq{eq_curly} is true for $n=k$. Differentiating the result once more with respect to $Q_{\nu_{k+1}}^{b_{k+1}}$ we obtain
\begin{align}   
    & gf^{eax}\left\{ \frac{\delta^{k-1}}{\delta Q_{\nu_{k+1}}^{b_{k+1}}(y_{k+1}) \delta Q_{\nu_k}^{b_k}(y_k) \cdots \delta Q_{\nu_3}^{b_{\s{3}}}(y_3)}  \left[Q_\mu^x(x)\Gamma_{Q_{\nu_1}^{b_{\s{1}}}Q_{\nu_2}^{b_{\s{2}}} Q_{\mu}^{e}}(y_1,y_2,x) \right]\right\}_{Q\rightarrow 0}= \nonumber\\
    &\hspace{3cm}
    gf^{eab_{\s{3}}}\delta(x-y_3)\Gamma_{Q_{\nu_1}^{b_{\s{1}}}Q_{\nu_2}^{b_{\s{2}}}Q_{\nu_3}^{e}Q_{\nu_4}^{b_4}\cdots Q_{\nu_k}^{b_k}Q_{\nu_{k+1}}^{b_{k+1}}}(y_1,y_2,x,y_4,\cdots,y_k,y_{k+1})  +  \cdots \nonumber \\
    &\hspace{3cm} + gf^{eab_k}\delta(x-y_k)\Gamma_{Q_{\nu_1}^{b_{\s{1}}}Q_{\nu_2}^{b_{\s{2}}}\cdots Q_{\nu_{k-1}}^{b_{k-1}}Q_{\nu_k}^{e}Q_{\nu_{k+1}}^{b_{k+1}}}(y_1,y_2,\cdots,y_{k-1},x,y_{k+1}) \nonumber \\ 
    &\hspace{3cm} +gf^{eab_{k+1}}\delta(x-y_{k+1})\Gamma_{Q_{\nu_1}^{b_{\s{1}}}Q_{\nu_2}^{b_{\s{2}}}\cdots Q_{\nu_{k}}^{b_{k}}Q_{\nu_{k+1}}^{e}}(y_1,y_2,\cdots,y_{k},x) \,. \nonumber \\
    &\hspace{3cm} + gf^{eax}\left[ Q_\mu^x(x)\Gamma_{Q_{\nu_1}^{b_{\s{1}}}Q_{\nu_2}^{b_{\s{2}}}\cdots Q_{\nu_{k}}^{b_{k}}Q_{\nu_{k+1}}^{b_{k+1}}Q_{\mu}^{e}}(y_1,y_2,\cdots,y_{k},y_{k+1},x) \right]_{Q\rightarrow0}   \,, 
\end{align}
which is \eq{eq_curly} for $n= k+1$. This completes the proof.

In momentum space, \eq{eq_bqn} is given by (suppressing a factor of $g$)
\begin{align}\label{finalsti}
    i q_\mu \Gamma_{B_\mu^aQ_{\nu_1}^{b_{\s{1}}}Q_{\nu_2}^{b_{\s{2}}}\cdots Q_{\nu_n}^{b_n}}(q,p_1,p_2, \cdots, p_n) = & f^{eab_{\s{1}}}  \Gamma_{Q_{\nu_1}^eQ_{\nu_2}^{b_{\s{2}}}Q_{\nu_3}^{b_{\s{3}}}\cdots Q_{\nu_n}^{b_n}}(p_1+q,p_2,p_3, \cdots, p_n)  \nonumber \\
    + & f^{eab_{\s{2}}}\Gamma_{Q_{\nu_1}^{b_{\s{1}}}Q_{\nu_2}^eQ_{\nu_3}^{b_{\s{3}}}\cdots Q_{\nu_n}^{b_n}}(p_1,p_2+q,p_3, \cdots, p_n)  +  \cdots  \nonumber \\
    + & f^{eab_n}\Gamma_{Q_{\nu_1}^{b_{\s{1}}}Q_{\nu_2}^{b_{\s{2}}}\cdots Q_{\nu_{n-1}}^{b_{n-1}}Q_{\nu_n}^{e}}(p_1,p_2,p_3, \cdots, p_n+q) \,.
\end{align}
The corresponding WI is obtained by expanding \1eq{finalsti} around $q = 0$ and collecting terms linear in $q$. Using   $p_n=-\sum\limits_{i=1}^{n-1} p_i$, we obtain 
\begin{align}
i \Gamma_{\!B_\mu^aQ_{\nu_1}^{b_{\s{1}}}Q_{\nu_2}^{b_{\s{2}}}\cdots Q_{\nu_n}^{b_n}}(0,p_1,p_2, \cdots, p_n) \!=\! 
\left(\!f^{eab_{\s{1}}}\frac{\partial}{\partial p_1^{\nu_1}} +\! \cdots \!+  f^{eab_{n-1}}\frac{\partial}{\partial p_{n-1}^{\nu_{n-1}}} \!\right)\Gamma_{\!Q_{\nu_1}^{b_{\s{1}}}\cdots Q_{\nu_n}^{e}}(p_1, \cdots, p_n) \,.
\end{align}  

Note that the absence of a zeroth order term on the l.h.s. 
of \1eq{finalsti} implies the relation 
\begin{align}
    0=&  f^{eab_{\s{1}}}  \Gamma_{Q_{\nu_1}^eQ_{\nu_2}^{b_{\s{2}}}Q_{\nu_3}^{b_{\s{3}}}\cdots Q_{\nu_n}^{b_n}}(p_1,p_2,p_3, \cdots, p_n)  
     + f^{eab_{\s{2}}}\Gamma_{Q_{\nu_1}^{b_{\s{1}}}Q_{\nu_2}^eQ_{\nu_3}^{b_{\s{3}}}\cdots Q_{\nu_n}^{b_n}}(p_1,p_2,p_3, \cdots, p_n) 
    +  \cdots \nonumber \\
    & \hspace{6cm}+  f^{eab_n}\Gamma_{Q_{\nu_1}^{b_{\s{1}}}Q_{\nu_2}^{b_{\s{2}}}\cdots Q_{\nu_{n-1}}^{b_{n-1}}Q_{\nu_n}^{e}}(p_1,p_2,p_3, \cdots, p_n) \,,
\label{zeroth}
\end{align}
whose validity we have checked explicitly for $n=3,4$. 

%---------------------------------------------------------------------------------%

\newpage 

\section{Feynman rules for BFM vertices} 
\label{sec:App_feynman}

In the Table~\ref{fig_feyback} of this Appendix we list the Feynman rules for BFM vertices at tree level.  
\begin{table}[ht]  
 \label{fig_feyback}
 \begin{tabular}{|m{0.124\textwidth}|c|} \hline
    % Head
    \parbox[c]{0.12\textwidth}{\textbf{Vertex}} & \textbf{Feynman rule} \\ \hline 
    %%%%%%%%%%%%%%%%%%%%%%%%%%%%%%%%%%%%%
    % First line - BQQ - fig10a
    %%%%%%%%%%%%%%%%%%%%%%%%%%%%%%%%%%%%%
    \parbox[c]{0.12\textwidth}{\vspace{0.1cm}\includegraphics[width=0.12\textwidth]{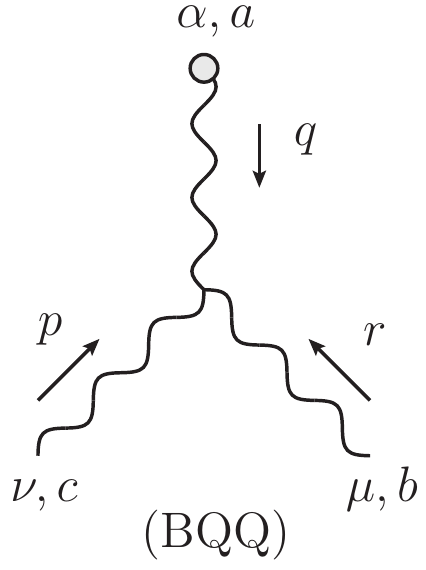}\vspace{0.1cm}} & 
    \begin{minipage}{14cm}\begin{eqnarray} \label{BQQ0}
        \gt_{\alpha\mu\nu}^{(0)}(q,r,p) =  (q-r)_{\nu}g_{\alpha\mu} &+& (r-p)_{\alpha} g_{\mu\nu} + (p-q)_{\mu}g_{\alpha\nu}  \\  &+& \xi_{\s Q}^{-1} (g_{\alpha\nu} r_\mu - g_{\alpha\mu} p_\nu) \,, \nonumber  
    \end{eqnarray} \end{minipage} \\ \hline
    %%%%%%%%%%%%%%%%%%%%%%%%%%%%%%%%%%%%%
    % Second line - Bcc - fig10b
    %%%%%%%%%%%%%%%%%%%%%%%%%%%%%%%%%%%%%
    \parbox[c]{0.12\textwidth}{\vspace{0.1cm}\includegraphics[width=0.12\textwidth]{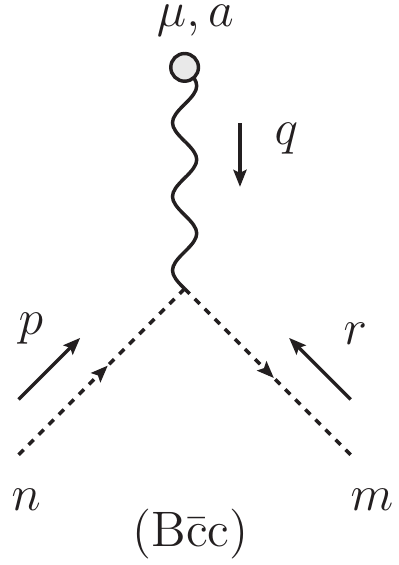}\vspace{0.1cm}} & 
    \begin{minipage}{14cm}\begin{eqnarray} \label{BCC0}
        \gt^{(0)}_\mu(r,p,q)=(r-p)_\mu \,, 
    \end{eqnarray} \end{minipage} \\ \hline
    %%%%%%%%%%%%%%%%%%%%%%%%%%%%%%%%%%%%%
    % Third line - BQQQ - fig10c
    %%%%%%%%%%%%%%%%%%%%%%%%%%%%%%%%%%%%%
    \parbox[c]{0.12\textwidth}{\vspace{0.1cm}\includegraphics[width=0.12\textwidth]{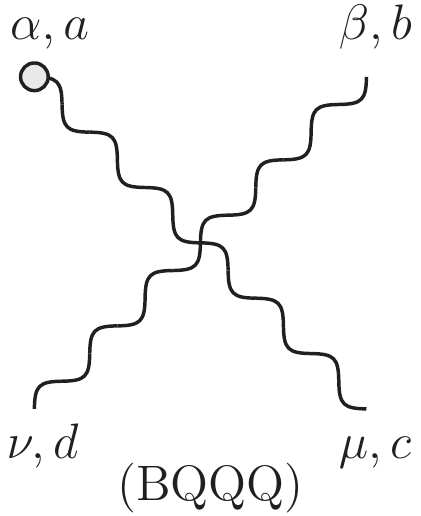}\vspace{0.1cm}} & 
    \begin{minipage}{14cm}\begin{eqnarray} \label{BQQQ0}
        \gt^{(0)abcd}_{\alpha\beta\mu\nu}=f^{adx}f^{cbx}\left(g_{\alpha\mu}g_{\beta\nu}-g_{\alpha\beta}g_{\mu\nu}\right)
		+f^{abx}f^{dcx}\left(g_{\alpha\nu}g_{\beta\mu}-g_{\alpha\mu}g_{\beta\nu}\right) \\
		+f^{acx}f^{dbx}\left(g_{\alpha\nu}g_{\beta\mu}-g_{\alpha\beta}g_{\mu\nu}\right) \,, \nonumber
    \end{eqnarray} \end{minipage} \\ \hline
    %%%%%%%%%%%%%%%%%%%%%%%%%%%%%%%%%%%%%
    % Forth line - BBQQ - fig10d
    %%%%%%%%%%%%%%%%%%%%%%%%%%%%%%%%%%%%%
    \parbox[c]{0.12\textwidth}{\vspace{0.1cm}\includegraphics[width=0.12\textwidth]{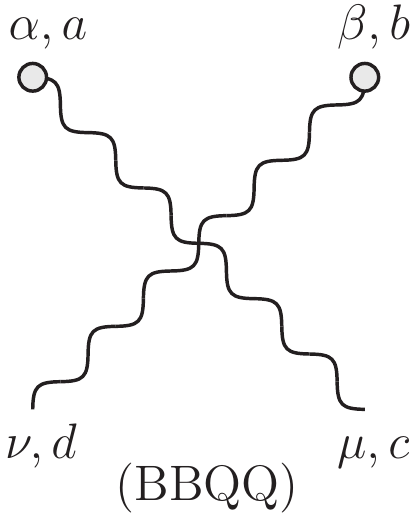}\vspace{0.1cm}} & 
    \begin{minipage}{14cm}\begin{eqnarray} \label{BBQQ0}
        \gh^{(0)abcd}_{\alpha\beta\mu\nu}&=\Gamma_{\alpha\beta\mu\nu}^{(0)abcd} + \xi_{\s Q}^{-1}(f^{adx}f^{bcx}g_{\alpha\nu}g_{\beta\mu} - f^{acx}f^{dbx}g_{\alpha\mu}g_{\beta\nu})\,,
    \end{eqnarray} \end{minipage} \\ \hline
    %%%%%%%%%%%%%%%%%%%%%%%%%%%%%%%%%%%%%
    % Fifth line - BQcc - fig10e
    %%%%%%%%%%%%%%%%%%%%%%%%%%%%%%%%%%%%%
    \parbox[c]{0.12\textwidth}{\vspace{0.1cm}\includegraphics[width=0.12\textwidth]{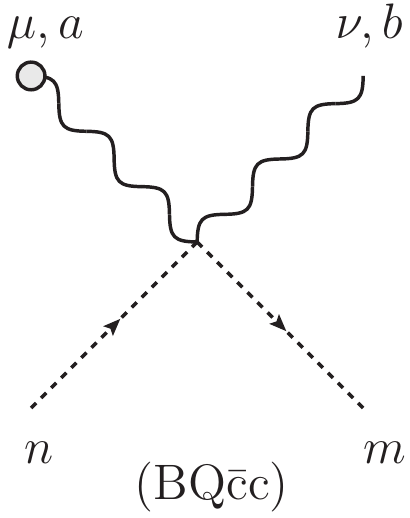}\vspace{0.1cm}} & 
    \begin{minipage}{14cm}\begin{eqnarray} \label{BQcc0}
        \gt^{(0)abmn}_{\mu\nu}=f^{max}f^{xbn}g_{\mu\nu} \,, 
    \end{eqnarray} \end{minipage} \\ \hline
    %%%%%%%%%%%%%%%%%%%%%%%%%%%%%%%%%%%%%
    % Sixth line - BBcc - fig10f
    %%%%%%%%%%%%%%%%%%%%%%%%%%%%%%%%%%%%%
    \parbox[c]{0.12\textwidth}{\vspace{0.1cm}\includegraphics[width=0.12\textwidth]{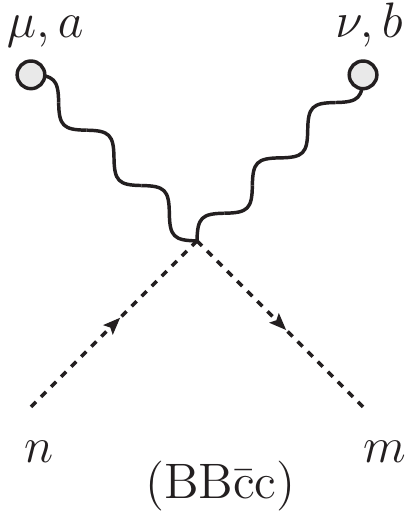}\vspace{0.1cm}} & 
    \begin{minipage}{14cm}\begin{eqnarray} \label{BBcc0}
        \gh^{(0)abmn}_{\mu\nu}=g_{\mu\nu}\left(f^{max}f^{bnx}+f^{mbx}f^{anx}\right) \,.
    \end{eqnarray} \end{minipage} \\ \hline
  \end{tabular}
  \caption{The diagrammatic representations of the new vertices appearing in the BFM and their respective Feynman rules at tree level~\cite{Binosi:2009qm}. Notice that for the three-point functions we have factored out the coupling $g$ and their respective color structure,  following the definitions of Eq.~\eqref{3vert}, while for the four-point functions,  we have factored out only $-ig^2$ as shown in Eq.~\eqref{4vert}.}
\end{table}

%%%%%%%%%%%%%%%%%%%%%%%%%%%%%%%%%%%%%

\newpage
\section{Abelian Slavnov-Taylor identities in the BFM} 
\label{apB}

In the Table~\ref{fig_wti} we collect all the Abelian STI in the BFM necessary to demonstrate the 
block-wise realization of the STI for the background three-gluon vertex.

\begin{table}[!htbp]
\setlength{\extrarowheight}{5pt}
\begin{tabularx}{\linewidth}{|c|X|}
\hline
% Head
\textbf{Vertex} & \hspace{5.5cm}\textbf{Abelian STI} \\
\hline
%%%%%%%%%%%%%
%First line - BQQ
%%%%%%%%%%%%%
BQQ & \vspace{-1.6cm} 
\begin{equation}
\label{BQQw}
    q^\alpha \, \gt_{\alpha\mu\nu}(q,r,p)  = \Delta^{-1}_{\mu\nu}(p) - \Delta^{-1}_{\mu\nu}(r) \,, \vspace{-0.9cm}
\end{equation}  \\ \hline
%%%%%%%%%%%%%
% Second line - Bcc
%%%%%%%%%%%%%
\bcc\ & \vspace{-1.6cm}
\begin{equation}
\label{BCCw}
    q^\mu \, \gt_\mu(r,p,q)  = D^{-1}(p) - D^{-1}(r) \,, \vspace{-0.9cm}
\end{equation} \\
\hline
%%%%%%%%%%%%%
% Third line - BQQQ
%%%%%%%%%%%%%
BQQQ & \vspace{-1.7cm}
\begin{eqnarray}    
\label{BQQQw}
        q^\alpha \gt^{abcd}_{\alpha\beta\mu\nu}(q,r,p,t) =& f^{abx}f^{dcx}\Gamma_{\beta\mu\nu}(r+q,p,t) +f^{acx}f^{bdx}\Gamma_{\beta\mu\nu}(r,p+q,t) \\
        &+f^{adx}f^{cbx}\Gamma_{\beta\mu\nu}(r,p,t+q)\,, \nonumber \vspace{-3cm}
\end{eqnarray} \vspace{-1cm}\\
\hline
%%%%%%%%%%%%%
% Forth line - BBQQ
%%%%%%%%%%%%%
BBQQ & \vspace{-1.7cm}
\begin{eqnarray}
\label{BBQQw}
    q^\alpha \gh^{abcd}_{\alpha\beta\mu\nu}(q,r,p,t) =&  f^{abx}f^{dcx}\gt_{\beta\mu\nu}(r+q,p,t) 
    +f^{acx}f^{bdx}\gt_{\beta\mu\nu}(r,p+q,t) \\ &+f^{adx}f^{cbx}\gt_{\beta\mu\nu}(r,p,t+q) \,, \nonumber \vspace{-7cm}
\end{eqnarray} \vspace{-1cm} \\
\hline
%%%%%%%%%%%%%
% Fifth line - BQcc
%%%%%%%%%%%%%
\bqcc\ \vspace{-0.2cm} & \vspace{-1.7cm}
\begin{eqnarray}
 \label{BQccw}
    q^\mu \gt^{abmn}_{\mu\nu}(q,r,p,t) =&f^{nax}f^{bmx}\Gamma_\nu(p,q+t,r) +f^{nbx}f^{max} \Gamma_\nu(q+p,t,r) \\
    &+f^{nmx}f^{abx}\Gamma_\nu(p,t,q+r) \nonumber  \,, \vspace{-5.5cm}
\end{eqnarray} \vspace{-0.6cm} \\
\hline
%%%%%%%%%%%%%
% Sixth line - BBcc
%%%%%%%%%%%%%
\bbcc & \vspace{-1.7cm}
\begin{eqnarray}
\label{BBccw}
    q^\mu \widehat{\Gamma}_{\mu\nu}^{abmn}(q,r,p,t) =& f^{abx}f^{mnx}\widetilde{\Gamma}_\nu(p,t,q+r) + f^{amx}f^{nbx}\widetilde{\Gamma}_\nu(q + p, t, r) \\   		 &+ f^{anx}f^{bmx}\widetilde{\Gamma}_\nu(p,q+t,r) \nonumber  \,, \vspace{-5.5cm}
\end{eqnarray} \vspace{-1cm}\\
\hline
%%%%%%%%%%%%%%%%
% Seventh line - BBQQQ
%%%%%%%%%%%%%%%%%
BBQQQ & \vspace{-1.7cm}
\begin{eqnarray}
\label{BBQQQw}
    q^\alpha\gh^{abcde}_{\alpha\beta\mu\nu\rho}(q,r,p,t,u) = f^{bax}\gt_{\beta\mu\nu\rho}^{xcde}(r+q,p,t,u) + f^{cax}\gt_{\beta\mu\nu\rho}^{bxde}(r,p+q,t,u)  \\
    + f^{dax}\gt_{\beta\mu\nu\rho}^{bcxe}(r,p,t+q,u) + f^{eax}\gt_{\beta\mu\nu\rho}^{bcdx}(r,p,t,u+q) \nonumber \,,
\end{eqnarray} \vspace{-1cm} \\
\hline
%%%%%%%%%%%%%%%%%
% Eighth line - BBQcc
%%%%%%%%%%%%%%%%%
\bbqcc\ & \vspace{-1.7cm}
\begin{eqnarray}
\label{BBQccw}
    q^\alpha \gh^{abcmn}_{\alpha\mu\nu}(q,r,p,t,u) =f^{bax}\gt_{\mu\nu}^{xcmn}(r+q,p,t,u)+ f^{cax}\gt_{\mu\nu}^{bxmn}(r,p+q,t,u)  \\            		+ f^{max}\gt_{\mu\nu}^{bcxn}(r,p,t+q,u)             		+ f^{nax}\gt_{\mu\nu}^{bcmx}(r,p,t,u+q) \nonumber \,.
\end{eqnarray} \vspace{-1cm} \\
\hline
\end{tabularx}
%%%%%%%%%%%%%%%%%%%%%%%%%%%%%%%%%%%%%%%%%
\caption{The Abelian STIs satisfied by the \bqq, \bcc, \bqqq, \bbqq, \bqcc, \bbcc, \bbqqq\ and \bbqcc\  vertices.}
	\label{fig_wti}
\end{table}

%%%%%%%%%%%%%%%%%%%%%%%%%%%%%%%
\newpage

\section{Expressions for the two-loop ghost sector of the \bbb\ SDE} 
\label{sec_expressions}

The two-loop ghost sector of the SDE of the vertex \bbb\, given by diagram $(e)$ in \fig{fig_bbbsde}, whose expansion is given in \fig{fig_2gh}, relate with the background gluon self-energy by \eq{eq_2gh}, where the expression for the diagrams  $(a_{\s{7}})$, $(a_{\s{8}})$, $(a_{\s{9}})$ and $(a_{\s{10}})$, in \fig{fig_bbsde} can be expressed~as
\begin{align} 
    (a_{\s{7}})^{ab}_{\alpha\mu}(q) &= ig^4 h_1^{aedm} \!\! \int_k \int_l  D(l) D(s) \Delta^{\beta}_\alpha (k)  \gt_{\mu\beta}^{bmde}(-q,k,l,s) \,, \label{eq_a7} \\
    (a_{\s{8}})^{ab}_{\alpha\mu}(q) &= \frac{1}{2}i\lambda^2 \delta^{ab} g_{\alpha\beta}  \!\! \int_k \int_l \yy_\sigma(-l,l-k) \yt_\mu^{\sigma \beta}(-k,q+k) \,, \label{eq_a8} \\
    (a_{\s{9}})^{ab}_{\alpha\mu}(q) &= -i\lambda^2\delta^{ab} \!\! \int_k \int_l D(q+l) \Delta_{\alpha}^{\beta}(k) \yy_\beta(k-l,l) \gt_\mu(-l,q+l,-q)\,, \label{eq_a9} \\
    (a_{\s{10}})^{ab}_{\alpha\mu}(q) &= -\frac{1}{2}i\lambda^2\delta^{ab} \!\! \int_k \int_l  D(q+l)  \Delta_{\alpha}^{\beta}(k) \yy_\beta(l,k-l)\gt_\mu(q+l,-l,-q) \,.  \label{eq_a10}
\end{align}
The decomposition of diagram $(e)$, given in \2eqs{eq_e}{edecomp}, can be separated in three groups: $(e_{1})^{abc}_{\alpha\mu\nu}$, with 
\begin{align}
    %
    % First group:    
    %
    (e_{1,1})^{abc}_{\alpha\mu\nu} &= ig^4 h_2^{amdce} g_{\alpha\beta} \!\! \int_k \int_l D(k) D(s)  \yt_\nu^{\beta\sigma}(l,-p-l) \gt^{bedm}_{\mu\sigma}(r,l+p,s,k)\,, \nonumber \\
    (e_{1,2})^{abc}_{\alpha\mu\nu} &= ig^4 h_2^{cmade} \!\! \int_k \int_l D(s) \Delta_{\alpha}^{\beta} (l) \yt_{\nu}(-p-k,k) \gt^{bedm}_{\mu\beta}(r,l,s,k+p)\,,\nonumber \\
    (e_{1,3})^{abc}_{\alpha\mu\nu} &= ig^4 h_2^{amecd}\!\! \int_k \int_l D(k)  \Delta_{\alpha}^{\beta} (l) \yt_{\nu}(k,-k-p) \gt^{bedm}_{\mu\beta}(r,l,k+p,s)\,, \nonumber \\
    (e_{1,4})^{abc}_{\alpha\mu\nu} &= ig^4 h_1^{amed} \!\! \int_k \int_l D(k) D(s) \Delta_{\alpha}^{\beta} (l) \gh_{\mu\nu\beta}^{bcedm}(r,p,l,s,k) \,,
\end{align}
$(e_{2})^{abc}_{\alpha\mu\nu}$, with 
\begin{align}
    %
    % Second group:
    %
    (e_{2,1})^{abc}_{\alpha\mu\nu} &= ig^4 h_2^{amebd} g_{\alpha\beta}\!\! \int_k \int_l D(k) D(s) \yt_\mu^{\beta\sigma}(l,-r-l) \gt_{\nu\sigma}^{cdem}(p,r+l,s,k)\,, \nonumber \\
    (e_{2,2})^{abc}_{\alpha\mu\nu} &= -\frac{i}{4}\lambda^2 f^{abc} \!\! \int_k \int_l \Delta^{\beta}_\alpha (l) \yy_{\rho}(s,k) \yt_{\mu}^{\sigma \rho}(p+l,q-l) \gt_{\nu\beta\sigma} (p,l,-p-l) \,, \nonumber \\
    (e_{2,3})^{abc}_{\alpha\mu\nu} &= -\frac{i}{4}\lambda^2  f^{abc}g_{\alpha\beta}  \!\! \int_k \int_l \Delta^{\lambda\rho}(q-l) \yy_{\lambda}(s,k) \yt_{\mu}^{\beta\sigma}(l,-r-l) \gt_{\nu\rho\sigma} (p,q-l,l+r) \,, \nonumber \\
    (e_{2,4})^{abc}_{\alpha\mu\nu} &= \frac{i}{4}\lambda^2  f^{abc} g_{\alpha\beta}  \!\! \int_k \int_l  D(s) \yt_\nu (-k-p,k) \yt_\mu^{\beta\sigma}(l,-r-l) \Gamma_{\sigma}(s,k+p,l+r) \,, \nonumber \\
    (e_{2,5})^{abc}_{\alpha\mu\nu} &= ig^4h_1^{xmbe}h_2^{amecx}g_{\alpha\beta} \!\! \int_k \int_l  D(s)  \yt_{\nu}(k,-k-p) \yt_\mu^{\beta\sigma}(l,-r-l) \Gamma_\sigma (k+p,s,l+r) \,, \nonumber \\
    (e_{2,6})^{abc}_{\alpha\mu\nu} &= \frac{i}{2}g^2 \lambda f^{aem} \!\! \int_k \int_l \Delta_{\alpha}^{\beta} (l) \Delta^{\rho\sigma}(q-l) \yy_{\rho}(s,k) \gh_{\mu\nu\beta\sigma}^{bcem}(r,p,l,q-l) \,,
\end{align}
and finally $(e_{3})^{abc}_{\alpha\mu\nu}$, with 
\begin{align}
    %
    % Third group:
    %
    (e_{3,1})^{abc}_{\alpha\mu\nu} &= ig^4 h_2^{bmade} \!\! \int_k \int_l D(s) \Delta_{\alpha}^{\beta} (l) \yt_{\mu}(-k-r,k) \gt^{cedm}_{\nu\beta}(p,l,s,k+r) \,, \nonumber \\
    (e_{3,2})^{abc}_{\alpha\mu\nu} &= ig^4 h_1^{xmce}h_2^{amebx} g_{\alpha\beta} \!\! \int_k \int_l D(k) \yt_\nu^{\beta\sigma}(l,-l-p) \yy_{\sigma}(s,k+r)  \gt_{\mu}(-k-r,k,r) \,, \nonumber \\
    (e_{3,3})^{abc}_{\alpha\mu\nu} &= -\frac{i}{4}\lambda^2 f^{abc}  \!\! \int_k \int_l  \Delta_{\alpha}^{\beta} (l)\yt_\nu (k,-k-p) \yy_{\beta}(k-q,s) \gt_{\mu}(k+p,q-k,r) \,,  \nonumber \\
    (e_{3,4})^{abc}_{\alpha\mu\nu} &= -\frac{i}{4}\lambda^2  f^{abc} \!\! \int_k \int_l \Delta_{\alpha}^{\beta}(l) \Gamma_{\beta}(s+p,k+r,l) \yt_{\mu}(-k-r,k) \yt_\nu (s,-s-p) \,, \nonumber \\
    (e_{3,5})^{abc}_{\alpha\mu\nu} &= -\frac{i}{2}\lambda^2 f^{abc} \!\! \int_k \int_l  \Delta_{\alpha}^{\beta} (l) \yy_{\beta}(s,k-q) \yt_{\mu}(-k-r,k) \gt_\nu (q-k,k+r,p) \,, \nonumber \\
    (e_{3,6})^{abc}_{\alpha\mu\nu} &= -\frac{i}{2}g^2\lambda f^{ade} \!\! \int_k \int_l D(k) \Delta_{\alpha}^{\beta} (l) \yy_{\beta}(k-q,s) \gh^{cbde}_{\mu\nu}(r,p,k,q-k) \,, \nonumber \\
    (e_{3,7})^{abc}_{\alpha\mu\nu} &= ig^4 h_2^{amedb}\!\! \int_k \int_l D(s) \Delta_{\alpha}^{\beta} (l) \yt_{\mu}(k,-k-r) \gt^{cedm}_{\nu\beta}(p,l,k+r,s) \,, \nonumber \\
    (e_{3,8})^{abc}_{\alpha\mu\nu} &= ig^4  h_1^{xmce}h_2^{amebx}  g_{\alpha\beta} \!\! \int_k \int_l  D(k) \yt_\nu^{\beta\sigma}(l,-l-p) \yy_{\sigma}(k+r,s) \gt_{\mu}(k,-k-r,r)  \,, \nonumber \\
    (e_{3,9})^{abc}_{\alpha\mu\nu} &= \frac{i}{2}\lambda^2 f^{abc}\!\! \int_k \int_l \Delta_{\alpha}^{\beta} (l)\yy_{\beta}(s,k-q) \gt_{\mu}(q-k,k+p,r) \yt_\nu (-p-k,k) \,, \nonumber \\
    (e_{3,10})^{abc}_{\alpha\mu\nu} &= \frac{i}{4}\lambda^2 f^{abc} \!\! \int_k \int_l \Delta_{\alpha}^{\beta} (l)  \Gamma_{\beta}(r-k,-t,l) \yt_{\mu}(-k,k-r)  \yt_\nu (t,q+k+l)\,, \nonumber \\
    (e_{3,11})^{abc}_{\alpha\mu\nu} &= \frac{i}{2}\lambda^2 f^{abc} \!\! \int_k \int_l \Delta_{\alpha}^{\beta} (l) \yy_{\beta}(k-q,s) \yt_{\mu}(k,-k-r) \gt_\nu (k+r,q-k,p) \,, \nonumber \\
    (e_{3,12})^{abc}_{\alpha\mu\nu} &= -ig^2\lambda f^{ade} \!\! \int_k \int_l D(k) \Delta_{\alpha}^{\beta} (l) \yy_{\beta}(s,k-q) \gh^{bcde}_{\mu\nu}(r,p,q-k,k) \,.
\end{align}

%\bibliography{bibliography_new}

%merlin.mbs apsrev4-1.bst 2010-07-25 4.21a (PWD, AO, DPC) hacked
%Control: key (0)
%Control: author (8) initials jnrlst
%Control: editor formatted (1) identically to author
%Control: production of article title (-1) disabled
%Control: page (0) single
%Control: year (1) truncated
%Control: production of eprint (-1) disabled
%

\end{document}